\documentclass[10pt]{iopart}
\pdfoutput=1 
\usepackage{graphicx} 
\usepackage{url}
\usepackage{algorithm}
\usepackage{algpseudocode}
\def\vector#1{\mbox{\boldmath $#1$}}

\begin{document}
\title{Multi-agent reinforcement learning using echo-state network and its application to pedestrian dynamics }

\author{Hisato Komatsu}

\address{Data Science and AI Innovation Research Promotion Center, Shiga University, 522-8522, Shiga, Japan }

\ead{hisato-komatsu@biwako.shiga-u.ac.jp}

\begin{abstract}
In recent years, simulations of pedestrians using the multi-agent reinforcement learning (MARL) have been studied.
This study considered the roads on a grid-world environment, and implemented pedestrians as MARL agents using an echo-state network and the least squares policy iteration method. Under this environment, the ability of these agents to learn to move forward by avoiding other agents was investigated. Specifically, we considered two types of tasks: the choice between a narrow direct route and a broad detour, and the bidirectional pedestrian flow in a corridor.
The simulations results indicated that the learning was successful when the density of the agents was not that high.
\end{abstract}


\maketitle

\section{Introduction }
Motions of groups of people or animals have been studied in many fields such as transportation engineering and active matter physics. Comprehensively understanding such motions through only experiments and observations is challenging. Thus, several studies have conducted computer simulations for a better understanding. Traditionally, animals (including humans) are assumed to obey certain mathematical rules in these simulations\cite{VCBjCS95,HM95,MIN99}. However, with the recent development of machine learning, simulation methods that reproduce animals by agents of reinforcement learning (RL) have been proposed\cite{MgLF14,MgLF17,ZL19,BI22,HLXFLY23}. RL in an environment with several agents exist is referred to as multi-agent reinforcement learning (MARL), and has been studied intensively to realize the competition or cooperation between agents. 

Currently, deep learning is usually used to implement RL agents, because it outperforms conventional methods of machine learning\cite{Mnih15,Mnih16,Schulman17}. However, training of the deep learning incurs high computational costs even in the case of a single-agent RL. Thus, algorithms with lower computational costs are useful, particularly when considering numerous of agents.

This study proposed an RL algorithm using an echo-state network (ESN). ESN is a type of reservoir computing that uses recurrent neural networks (RNN) and trains only the output weight matrix. This restriction reduced the computational cost compared to that of deep learning, which trains all parameters of the neural network. Several methods have been proposed to implement RL using ESN\cite{SGL06,OKKhP11,CF20,ZLSZ21,CSYZHL18}. Among these methods, this study adopted a method similar to, but simpler than that of Ref.~\cite{ZLSZ21}. As a specific task, we considered pedestrians proceeding along a road. All pedestrians were RL agents that received positive(negative) rewards when they travelled along the same(opposite) direction as their targeted direction. Although the task itself was simple, other agents were employed as the obstacles preventing each agent's walking. Hence, the problem increased in complexity with increasing number of agents. We investigated whether the agents could proceed in the direction they wanted to by avoiding other agents. 

The remainder of this paper is organized as follows: first, we review the related previous studies in Section \ref{related_work}. Then, our method is proposed in Section \ref{method}, and the results of numerical simulations are shown in Section \ref{results}.
Finally, we summarize the study in Section \ref{summary}.


\section{Related work \label{related_work} }

\subsection{Multi-agent reinforcement learning (MARL) }

In environments with several agents, RL agents are required to cooperate or compete with other agents. Such environments appear in various fields, thus, MARL has been studied intensively\cite{BBdS08}. In particular, methods of deep reinforcement learning (DRL) have been applied to MARL, as well as single-agent RL\cite{DD21,GD22,OH23}.

The simplest method to implement MARL is to allow the agents to learn their policies independently. However, independent learners cannot access the experiences of other agents, and consider other agents as a part of environment. This results in the non-stationarity of the environment from the perspective of each agent. To address this problem, many algorithms have been proposed. For example, the multi-agent deep deterministic policy gradient (MADDPG) algorithm adopts the actor-critic method, and allows the critic to observe all agents' observations, actions, and target policies\cite{LWTHAM17}. This algorithm was improved and applied to the pedestrian dynamics by Zheng and Liu\cite{ZL19}. Gupta et al. introduced a parameter sharing method wherein agents shared the parameters of their neural networks\cite{GEK17}. Their method exhibited high learning efficiency in environments with homogeneous agents.

This study adopted the parameter sharing method because it can be easily extended to a form suitable to ESN. Specifically, we considered one or two groups of agents that shared their experiences and policies but received rewards separately.

\subsection{Echo-state network (ESN) \label{review_ESN} }
As mentioned earlier, ESN is a type of reservoir computing that uses RNN\cite{Jaeger01,Lukosevicius12,ZV23}:
\begin{eqnarray}
\tilde{ \vector{X} }^{\mathrm{res} } (t) & = & f \left( \vector{W} ^{\mathrm{in} } _s \vector{X} ^{\mathrm{in} } (t) + \vector{W} ^{\mathrm{in} } _b + \vector{W} ^{\mathrm{res} } \vector{X} ^{\mathrm{res} } (t-1) \right) , \\
\vector{X} ^{\mathrm{res} } (t) & = & \alpha \tilde{ \vector{X} }^{\mathrm{res} } (t) + (1 - \alpha)  \vector{X} ^{\mathrm{res} } (t-1) , \\
\vector{X} ^{\mathrm{out} } (t) & = & \vector{W} ^{\mathrm{out} } \left( 
\begin{array}{c}
 \vector{X} ^{\mathrm{res} } (t) \\
 1
 \end{array}
 \right) ,
\end{eqnarray}
where $\vector{X} ^{\mathrm{in} } (t)$ and $\vector{X} ^{\mathrm{out} } (t)$ are the input and output vectors at time $t$, respectively, $f$ is the activation function, and $\alpha$ is the constant called leaking rate. In the case of ESN, elements of matrices $\vector{W} ^{\mathrm{in} } _s$, $\vector{W} ^{\mathrm{in} } _b$, and $\vector{W} ^{\mathrm{res} }$ are randomly fixed, and only the output weight matrix $\vector{W} ^{\mathrm{out} }$ is trained. Through this restriction of the trainable parameters, the computational cost is considerably lowered compared with the deep learning. In addition, the exploding/vanishing gradient problem does not occur unlike the training of RNN using backpropagation. For example, in the case of supervised learning, training is typically conducted by the Ridge regression that minimizes the mean square error between the output vector $\vector{X} ^{\mathrm{out} }$ and the training data $\vector{X} ^{\mathrm{train} }$. This calculation is considerably faster than the backpropagation used in the deep learning. 

However, in the case of RL the training data should be collected by agents themselves. Hence, the training method is not as easy as in the case of the supervised learning. Many previous studies have proposed different algorithms to realize RL using ESN. For example, Szita et al. used the SARSA method\cite{SGL06}, Oubbati et al. adopted ESN as a critic of the actor-critic method\cite{OKKhP11}, Chang and Futagami used a covariance matrix adaptation evolution strategy (CMA-ES)\cite{CF20}, and Zhang et al. improved the least squares policy iteration (LSPI) method\cite{ZLSZ21}. Previous studies have also utilized ESN to MARL. For example, Chang et al. trained ESN employing a method similar to that of a deep Q network(DQN), and applied it to the distribution of the radio spectra of telecommunication equipment\cite{CSYZHL18}. 
In this study, we adopted the LSPI method, as in Ref.~\cite{ZLSZ21}; however, we used a simpler algorithm. Note that the absence of the exploding/vanishing gradient problem in the ESN comes from the point that the weight matrices involved in this problem are fixed. Hence, although the above-mentioned training algorithms including LSPI are more complex than those of supervised learning, such problem still does not appear.

Typically, the recurrent weight $\vector{W} ^{\mathrm{res} }$ is a sparse matrix with spectral radius $\rho < 1$. To generate such a matrix, first, a random matrix $\vector{W} _0 ^{\mathrm{res} }$ must be created whose components $W _{0, \mu \nu} ^{\mathrm{res} }$ are expressed as:
\begin{equation}
\left\{ 
\begin{array}{cc}
W _{0, \mu \nu} ^{\mathrm{res} } = 0 & \mathrm{with \ prob. \ } p_s ^{\mathrm{res} } \\
W _{0, \mu \nu} ^{\mathrm{res} } \sim \mathcal{P ^{\mathrm{res} } } & \mathrm{with \ prob. \ } 1-p_s ^{\mathrm{res} } 
\end{array}
 \right. ,
\end{equation}
where $\mathcal{P ^{\mathrm{res} } }$ is a predetermined probability distribution and $p_s ^{\mathrm{res} }$ is a hyperparameter called sparsity. Regularizing this matrix such that its spectral radius is expressed as $\rho$, we obtain $\vector{W} ^{\mathrm{res} }$:
\begin{equation}
\vector{W} ^{\mathrm{res} } = \frac{\rho}{\rho \left( \vector{W} _0 ^{\mathrm{res} } \right) } \vector{W} _0 ^{\mathrm{res} } .
\end{equation}
where $\rho \left( \vector{W} _0 ^{\mathrm{res} } \right) $ is the spectral radius of $\vector{W} _0 ^{\mathrm{res} }$. According to Ref.~\cite{Lukosevicius12}, the input weights $\vector{W} ^{\mathrm{in} } _s$ and $\vector{W} ^{\mathrm{in} } _b$ are usually generated as dense matrices. However, we also rendered them sparse, as has been explained later.

\subsection{Least squares policy iteration (LSPI) method \label{review_LSPI} }
The LSPI method is an RL algorithm that was proposed by Lagoudakis and Parr\cite{LP03}. It approximates the state-action value function $Q(s,a)$ as the linear combination of given functions of $s$ and $a$, $\vector{\phi} (s,a) = \left( \phi _1 (s,a) , ... , \phi _N (s,a) \right) ^T $:
\begin{equation}
Q(s,a) = \sum _{i=1} ^N w _i \phi _i (s,a) = \vector{w}^T \vector{\phi} (s,a) , \label{Q_LSPI}
\end{equation}
where $\vector{w} = (w_1 , ... , w_N )^T$ is the coefficient vector that is to be trained. Using this value, action $a_t$ of the agent at each time step $t$ is decided by the $\epsilon$-greedy method:
\begin{equation}
a_t = \left\{ 
\begin{array}{cc}
\mathrm{argmax} _a Q(s_t , a) & \mathrm{with \ prob. \ } 1-\epsilon \\
\mathrm{randomly \ chosen \ action } & \mathrm{with \ prob. \ } \epsilon
\end{array}
 \right. , \label{epsilon_greedy}
\end{equation}
where $s_t$ is the state at this step. Usually, $\epsilon$ is initially set as a relatively large value $\epsilon _0$, and gradually decreases.

To search the appropriate policy, the LSPI method updates $\vector{w}$ iteratively. Let the $n$-th update be executed after the duration $\mathcal{E} _n$. Then the form of \vector{w} after the $n_l$-th duration, $\mathcal{E} _{n_l}$, is expressed as:
\begin{equation}
\vector{w} = A^{-1} B , \label{w_LSPI}
\end{equation}
where 
\begin{eqnarray}
A & = & \sum _{n=1} ^{n_l} \lambda^{n_{l} -n} \sum _{t \in \mathcal{E}_n } \vector{\phi} (s_t , a_t ) \left( \vector{\phi} (s_t , a_t ) - \gamma \vector{\phi} (s_{t+1} , a_{t+1} ) \right)^T , \\
B & = & \sum _{n=1} ^{n_l} \lambda^{n_{l} -n} \sum _{t \in \mathcal{E}_n } r_t \vector{\phi} (s_t , a_t ) ,
\end{eqnarray}
$r_t$ is the reward at this step, and $\gamma \in (0,1)$ is the discount factor. We let $\vector{\phi} (s_{t+1} , a_{t+1} ) = 0$ if the episode ends at time $t$. The constant $\lambda \in (0,1)$ is a forgetting factor that is introduced because the contributions of old experiences collected under the old policy do not reflect the response of the environment under the present policy. This factor has also been adopted by previous studies such as Ref.~\cite{ZLSZ21}.

\section{Proposed method and settings of simulations \label{method} }

\subsection{Application of ESN to the LSPI method \label{ESN_LSPI} }

In this study, we considered the partially observable Markov decision process (POMDP), a process wherein each agent observed only a part of the state. As the input of the $i$-th agent, we set its observation $\vector{o}_{i,t}$, the candidate of the action $a$, and the bias term:
\begin{eqnarray}
\tilde{ \vector{X} } _i ^{\mathrm{res} } (t;a) & = & f \left( \vector{W} ^{\mathrm{in} } _o \vector{o}_{i;t}  + \vector{W} ^{\mathrm{in} } _a \vector{\eta} _a + \vector{W} ^{\mathrm{in} } _b + \vector{W} ^{\mathrm{res} } \vector{X} _i ^{\mathrm{res} } (t-1) \right) , \label{tildeXres_proposed} \\
\bar{ \vector{X} } _i ^{\mathrm{res} } (t;a) & = & \alpha \tilde{ \vector{X} } _i ^{\mathrm{res} } (t;a) + (1 - \alpha)  \vector{X} _i ^{\mathrm{res} } (t-1) , \label{Xres_proposed} \\
Q_i (\vector{o}_{i;t},a) & = & \vector{W} _i ^{\mathrm{out} } \left( 
\begin{array}{c}
 \bar{ \vector{X} } _i ^{\mathrm{res} } (t;a) \\
 1
 \end{array}
 \right) , \label{Q_proposed}
\end{eqnarray}
where $\vector{\eta}_a$ is the one-hot expression of $a$ and activation function $f$ is ReLU. We expressed the number of neurons of the reservoir as $N ^{\mathrm{res} }$, i.e. $N ^{\mathrm{res} } \equiv \mathrm{dim} \vector{X} _i ^{\mathrm{res} } $. The initial condition of reservoir was set as $\vector{X} _i ^{\mathrm{res} } (-1) = 0$. We assumed that all agents shared the common input and reservoir matrices $\vector{W} ^{\mathrm{in} } _o $, $\vector{W} ^{\mathrm{in} } _a $, $\vector{W} ^{\mathrm{in} } _b $, and $\vector{W} ^{\mathrm{res} } $. For later convenience, $\vector{W} ^{\mathrm{in} } _o$, $\vector{W} ^{\mathrm{in} } _a $, and $\vector{W} ^{\mathrm{in} } _b $, that is, the parts of the input matrix are denoted separately.
The action $a_t$ was decided by the $\epsilon$-greedy method, Eq. (\ref{epsilon_greedy}), under this $Q_i (\vector{o}_{i;t},a)$. Here, $Q_i$ should be calculated for all possible candidate of action $a$. Using the accepted action, $\vector{X} _i ^{\mathrm{res} }$ is updated as follows:
\begin{equation}
\vector{X} _i ^{\mathrm{res} } (t) =\bar{ \vector{X} } ^{\mathrm{res} } (t;a_t) \label{X_chosen}
\end{equation} 
In the actual calculation of Eq.~(\ref{tildeXres_proposed}), it is convenient to calculate $\tilde{ \vector{X} } _i ^{\mathrm{res} } (t;a) $ for all candidates of action $a$ at once using the following relation:
\begin{eqnarray}
& & \left( \tilde{ \vector{X} } _i ^{\mathrm{res} } (t;a=1) , \cdots \tilde{ \vector{X} } _i ^{\mathrm{res} } (t;a=|\mathcal{A}| ) \right) \nonumber \\
& = & f \left(  \left( \vector{X} _t ^{\mathrm{in} } + \vector{W} ^{\mathrm{in} } _a \vector{\eta} _{a=1} , \cdots , \vector{X} _t ^{\mathrm{in} } + \vector{W} ^{\mathrm{in} } _a \vector{\eta} _{a=|\mathcal{A}| } \right) \right) \nonumber \\
 & = & f \left( \mathrm{rep} _{|\mathcal{A} | } \left( \vector{X} _t ^{\mathrm{in} } \right) + \vector{W} ^{\mathrm{in} } _a \right) , \label{tildeXres_matrix}
\end{eqnarray}
\begin{equation}
\mathrm{where} \ \vector{X} _t ^{\mathrm{in} } \equiv \vector{W} ^{\mathrm{in} } _o \vector{o}_{i;t}  +  \vector{W} ^{\mathrm{in} } _b + \vector{W} ^{\mathrm{res} } \vector{X} _i ^{\mathrm{res} } (t-1) . \label{Xin_def}
\end{equation}
Here, $|\mathcal{A}| $ is the number of candidates of the action, and $\mathrm{rep} _{|\mathcal{A} | } \left( \vector{X} _t ^{\mathrm{in} } \right)$ is a matrix constructed by repeating the column vector, $\vector{X} _t ^{\mathrm{in} }$, $|\mathcal{A}| $-times. To derive the last line of Eq.~(\ref{tildeXres_matrix}), we should remind that $\vector{\eta} _a$ is the one-hot expression:
\begin{equation}
\vector{W} ^{\mathrm{in} } _a \left( \vector{\eta} _{a=1}, \cdots , \vector{\eta} _{a=|\mathcal{A} | } \right) = \vector{W} ^{\mathrm{in} } _a \left( 
\begin{array}{cccc}
1 & 0 & \cdots & 0 \\
0 & 1 & \cdots & 0 \\
& & \vdots & \\
0 & 0 & \cdots & 1 \\
\end{array}
\right) = \vector{W} ^{\mathrm{in} } _a .
\end{equation}

Comparing Eqs.~(\ref{Q_LSPI}), (\ref{Q_proposed}) and (\ref{X_chosen}), we can apply the LSPI method by substituting $\vector{w} ^T = \vector{W} _i ^{\mathrm{out} } $ and $\vector{\phi} = \left( \vector{X} _i ^{\mathrm{res} } (t) ^T , 1\right) ^T$. Thus, taking the transpose of Eq.(\ref{w_LSPI}), $\vector{W} _i ^{\mathrm{out} }$ can be calculated by the following relation
\begin{equation}
\vector{W} _i ^{\mathrm{out} } = \tilde{B}_{G_i} \tilde{A} _{G_i} ^{-1} , \label{Wout_ESN}
\end{equation}
where 
\begin{eqnarray}
\tilde{A} _{G_i} & = & \sum _{n=1} ^{n_l} \lambda^{n_{l} -n} \sum _{j \in G_i} \sum _{t \in \mathcal{E}_n }  \left( \hat{ \vector{X} } _j ^{\mathrm{res} } (t) - \gamma \hat{ \vector{X} } _j ^{\mathrm{res} } (t+1) \right) \hat{ \vector{X} } _j ^{\mathrm{res} } (t) ^T \nonumber \\
& & + \lambda ^{n_l - 1} \beta I \label{A_ESN} \\
\tilde{B} _{G_i} & = & \sum _{n=1} ^{n_l} \lambda^{n_{l} -n} \sum _{j \in G_i} \sum _{t \in \mathcal{E}_n } r_{j;t} \hat{ \vector{X} } _j ^{\mathrm{res} } (t) ^T , \label{B_ESN} 
\end{eqnarray}
and
\begin{equation}
\hat{ \vector{X} } _j ^{\mathrm{res} } (t)  \equiv  \left(
\begin{array}{c}
\vector{X} _i ^{\mathrm{res} } (t) \\
 1
 \end{array}
 \right)  . 
\end{equation}
Here, we divided the agents into one or several groups, and let all members of the same group share their experiences when we calculated $\vector{W} _i ^{\mathrm{out} } $. In Eqs.~(\ref{Wout_ESN}), (\ref{A_ESN}) and (\ref{B_ESN}), $G_i$ indicates the group to which the $i$-th agent belongs. In Eq.~(\ref{A_ESN}), we let $\hat{ \vector{X} } _j ^{\mathrm{res} } (t+1) = 0$ if the episode ended at time $t$, as in Section \ref{review_LSPI}. The second term of the right-hand side of Eq.~(\ref{A_ESN}) was introduced to avoid the divergence of the inverse matrix of Eq.~(\ref{Wout_ESN}). The constant $\beta$ is a small hyperparameter, and $I$ is the identity matrix. As the duration $\mathcal{E}_n$, we adopted one episode in this study. In other words, the output weight matrix $\vector{W} _i ^{\mathrm{out} }$ was updated at the end of each episode. The agents of the same group shared the common output weight matrix by Eq.~(\ref{Wout_ESN}). Considering that all agents had the same input and reservoir matrices, as explained above, agents of the same group shared all parameters of neural networks. Thus, this is a form of parameter sharing suitable for ESN. 
As mentioned in Section \ref{review_ESN}, Ref.~\cite{ZLSZ21} also applied LSPI method to train ESN. They calculated the output weight matrix using the recursive least squares(RLS) method to avoid the calculation of the inverse matrix of Eq.~(\ref{Wout_ESN}), which incurred a computational cost of $O \left( (N ^{\mathrm{res} })^3 \right)$. However, we did not adopt RLS method, because an additional approximation called mean-value approximation is required for this method. In addition, when each duration $\mathcal{E} _n$ is long and the parameter sharing method is introduced, the effect of the computational cost mentioned above on the total computation time is limited. 

In the actual calculation, we made the matrices
\begin{eqnarray}
Y_1 \equiv \left( 
\begin{array}{c}
\left( \hat{ \vector{X} } _{j = 1} ^{\mathrm{res} } (0) - \gamma \hat{ \vector{X} } _{j = 1} ^{\mathrm{res} } (1) \right) ^T \\
\vdots \\
\left( \hat{ \vector{X} } _{j = |G| } ^{\mathrm{res} } (0) - \gamma \hat{ \vector{X} } _{j = |G|} ^{\mathrm{res} } (1) \right) ^T \\
\vdots \\
\left( \hat{ \vector{X} } _{j = 1} ^{\mathrm{res} T} (t_{\mathrm{max} } - 1 ) - \gamma \hat{ \vector{X} } _{j = 1} ^{\mathrm{res} } (t_{\mathrm{max} } ) \right) ^T \\
\vdots \\
\left( \hat{ \vector{X} } _{j = |G| } ^{\mathrm{res} T} (t_{\mathrm{max} } - 1 ) - \gamma \hat{ \vector{X} } _{j = |G|} ^{\mathrm{res} T} (t_{\mathrm{max} } ) \right) ^T \\
\left( \hat{ \vector{X} } _{j = 1} ^{\mathrm{res} } (t_{\mathrm{max} } ) \right) ^T \\
\vdots \\
\left( \hat{ \vector{X} } _{j = |G| } ^{\mathrm{res} } (t_{\mathrm{max} } ) \right) ^T \\
 \end{array}
\right) , Y_2 \equiv \left( 
\begin{array}{c}
\left( \hat{ \vector{X} } _{j = 1} ^{\mathrm{res} } (0) \right) ^T \\
\vdots \\
\left( \hat{ \vector{X} } _{j = |G| } ^{\mathrm{res} } (0) \right) ^T \\
\vdots \\
\left( \hat{ \vector{X} } _{j = 1} ^{\mathrm{res} } (t_{\mathrm{max} } - 1 ) \right) ^T \\
\vdots \\
\left( \hat{ \vector{X} } _{j = |G| } ^{\mathrm{res} } (t_{\mathrm{max} } - 1 ) \right) ^T \\
\left( \hat{ \vector{X} } _{j = 1} ^{\mathrm{res} } (t_{\mathrm{max} } ) \right) ^T \\
\vdots \\
\left( \hat{ \vector{X} } _{j = |G| } ^{\mathrm{res} } (t_{\mathrm{max} } ) \right) ^T \\
 \end{array}
\right) , \nonumber \\
\label{Y_matrix} 
\end{eqnarray}
and
\begin{eqnarray}
R \equiv \left( r _{j = 1 ; t=0}, \cdots , r _{j = |G| ; t=0} , \cdots, r _{j = |G| ; t=t_{\mathrm{max} } -1 } , \cdots, r _{j = |G| ; t=t_{\mathrm{max} } -1 }, 0, \cdots , 0 \right)^T , \nonumber \\
\label{R_matrix} 
\end{eqnarray}
, and performed the operation 
\begin{eqnarray}
\tilde{A} _{G} & \leftarrow & \tilde{A} _{G} + Y_1 ^T Y_2 , \label{A_ESN_each_episode} \\
\tilde{B} _{G} & \leftarrow & \tilde{B} _{G} + R ^T Y_2 , \label{B_ESN_each_episode} 
\end{eqnarray}
at the end of each episode $\mathcal{E}_n$. Here, we used the fact that the reward after the end of the episode is zero: $r _{j;t_{\mathrm{max}} } = 0$. However, the term $\hat{ \vector{X} } _j ^{\mathrm{res} } (t_{\mathrm{max} }) \neq 0$ should not be ignored. After this manipulation, we calculated $\vector{W} _i ^{\mathrm{out} }$ using Eq. (\ref{Wout_ESN}), multiplying forgetting factors to matrices $\tilde{A} _{G}$ and $\tilde{B} _{G}$:
\begin{eqnarray}
\tilde{A} _{G} & \leftarrow & \lambda \tilde{A} _{G} ,\\
\tilde{B} _{G} & \leftarrow & \lambda \tilde{B} _{G} ,
\end{eqnarray}
 and updated the value of $\epsilon$ used for the $\epsilon$-greedy method as:
\begin{equation}
\epsilon \leftarrow \delta _{\epsilon} \epsilon ,
\end{equation}
if $\epsilon$ was larger than the predetermined threshold value $\epsilon _{\mathrm{min} } $. Training is dependent on the information in past state, action, and reward only through Eqs. (\ref{A_ESN_each_episode}) and (\ref{B_ESN_each_episode}). Thus, matrices $Y_1, Y_2,$ and $R$ can be deleted after the update of $\tilde{A} _{G}$ and $\tilde{B} _{G}$ using these equations.
Note that, in principle, we can update $\tilde{A} _{G}$ and $\tilde{B} _{G}$ by adding the increments every time step even if we do not prepare matrices $Y_1, Y_2,$ and $R$. However, the number of calculations of large matrices multiplication should be reduced to lower the computational cost, at least when the calculation is implemented by Python. 
The algorithm explained above is summarized in Algorithm \ref{alg1}. 

\begin{figure}[!t]
\begin{algorithm}[H]
    \caption{LSPI algorithm used in this study}
    \label{alg1}
\begin{algorithmic}[1]
   \State $\epsilon \gets \epsilon _0$
   \ForAll {$G$ : group} 
      \State $\tilde{A}_G \gets \beta I $
   \EndFor
   \For {$n \gets 1, ... , n_{\mathrm{max} } $}
      \For {$t \gets 0, ... , t_{\mathrm{max} } -1 $} 
         \State decide each agent's action 
         \State update the environment
         \State give agents rewards $r_{j;t}$
      \EndFor
      \ForAll {$G$ : group }
         \State $\tilde{A} _{G} \leftarrow \tilde{A} _{G} + Y_1 ^T Y_2 $
         \State $\tilde{B} _{G} \leftarrow \tilde{B} _{G} + R ^T Y_2 $ 
         \ForAll {$j \in G $} 
            \State $\vector{W} _j ^{\mathrm{out} } \gets \tilde{B}_G \tilde{A}_G ^{-1} $
         \EndFor
         \State $\tilde{A}_G \gets \lambda \tilde{A}_G$
         \State $\tilde{B}_G \gets \lambda \tilde{B}_G$
      \EndFor
      \If {$\epsilon > \epsilon _{\mathrm{min} }$}
         \State $\epsilon \gets \delta _{\epsilon} \epsilon$
      \EndIf 
      \State reset the environment
   \EndFor
\end{algorithmic}
\end{algorithm}
\end{figure}

\subsection{environment and observation of each agent}
\begin{figure}[!tb]
 \centering
\includegraphics[width = 6.0cm]{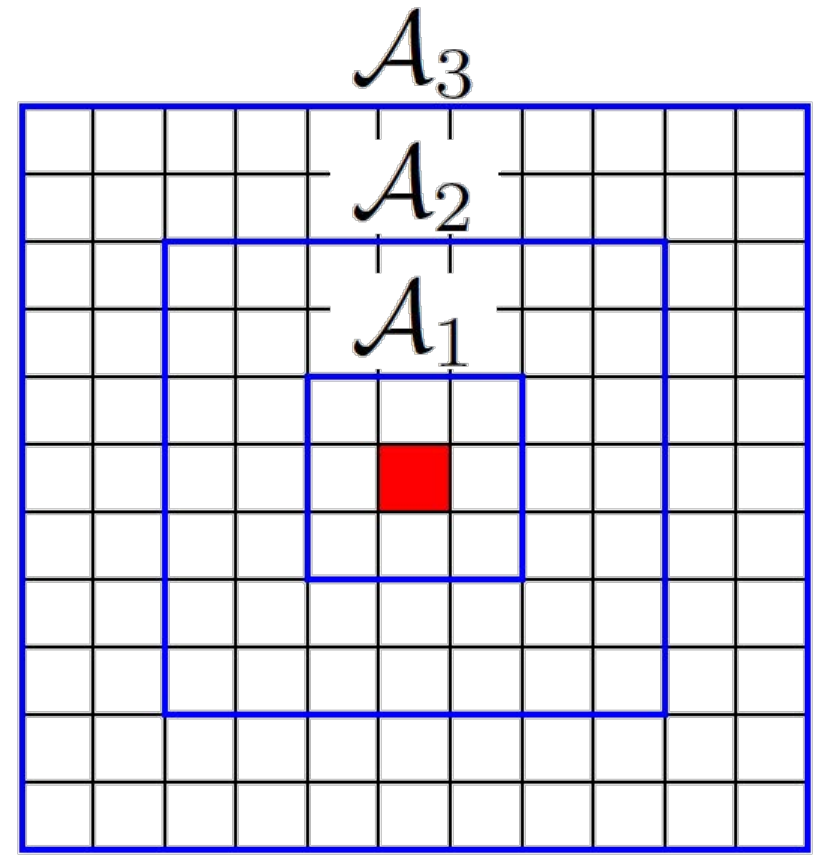}
\caption{Eyesight of each agent. We called $3 \times 3$, $7 \times 7$, and $11 \times 11$ cells centering on agent itself (painted red in this figure) as $\mathcal{A}_{1}$, $\mathcal{A}_{2}$, and $\mathcal{A}_{3}$, respectively. }
\label{eyesight}
\end{figure}

We constructed a grid-world environment wherein each cell could have three states: ``vacant,'' ``wall,'' and ``occupied by an agent.'' At each step, each agent chose one of the four adjacent cells: up, down, right, and left, and attempted to move into it. This move succeeded if and only if the candidate cell was vacant and no other agents attempted to move there. Each agent observed $11 \times 11$ cells centering on the agent itself as 2-channel bitmap data. Here, agents including observer itself are indicated as pairs of numbers $(1,0)$, walls are as $(0,1)$, and vacant cells are as $(0,0)$. Hence, the observation of $i$-th agent, $\vector{o} _{i;t} $, was $11 \times 11 \times 2 = 242$-dimensional vector. Although several previous studies have proposed a method that combined ESN with untrained convolutional neural networks when it is used for image recognition\cite{CF20,TT18,TT22}, we input the observation directly into ESN.
Instead, we improved the sparsity of input weight matrix $\vector{W} ^{\mathrm{in} } _o$ to reflect the spatial structure. Specifically, we first called the $3 \times 3$, $7 \times 7$, and $11 \times 11$ cells centering on agent as $\mathcal{A}_{1}$, $\mathcal{A}_{2}$ and $\mathcal{A}_{3}$ respectively, as in Fig.~\ref{eyesight}. Then, we changed the sparsity of the components of $\vector{W} ^{\mathrm{in} } _o$ depending on which cell they combined with. Namely, components of $W _{o,\mu \nu} ^{\mathrm{in} }$ are generated as:
\begin{equation}
\left\{ 
\begin{array}{cc}
W _{o,\mu \nu} ^{\mathrm{in} } = 0 & \mathrm{with \ prob. \ } p_s ^{\mathrm{in} } ( \nu ) \\
W _{o,\mu \nu} ^{\mathrm{in} } \sim \mathcal{P} ^{\mathrm{in} } _{o} & \mathrm{with \ prob. \ } 1-p_s ^{\mathrm{in} } ( \nu )
\end{array}
 \right. ,
\end{equation}
and the sparsity $p_s ^{\mathrm{in} } ( \nu )$ is expressed as follows:
\begin{equation}
p_s ^{\mathrm{in} } ( \nu ) = \left\{ 
\begin{array}{cc}
p_{s1} ^{\mathrm{in} } & \mathrm{if} \ \nu \in \mathcal{A}_1 \\
p_{s2} ^{\mathrm{in} } & \mathrm{if} \ \nu \in \mathcal{A}_2 \backslash \mathcal{A}_1 \\
p_{s3} ^{\mathrm{in} } & \mathrm{if} \ \nu \in \mathcal{A}_3 \backslash \mathcal{A}_2
\end{array}
 \right. ,
\end{equation}
where the hyperparameters $p_{s1} ^{\mathrm{in} } $, $p_{s2} ^{\mathrm{in} } $, and $p_{s3} ^{\mathrm{in} } $ obey $p_{s1} ^{\mathrm{in} } \leq p_{s2} ^{\mathrm{in} } \leq p_{s3} ^{\mathrm{in} }$. Thus, information on cells near the agent combined with the reservoir more densely than that on cells far from the agent. Further, we let $\vector{W} ^{\mathrm{in} } _b $ be a sparse matrix with sparsity $p_{sb} ^{\mathrm{in} }$, but made $\vector{W} ^{\mathrm{in} } _a $ dense. In addition, $\vector{W} ^{\mathrm{res} } $ was generated as a sparce matrix with its spectral radius $\rho < 1$, by the process explained in Section \ref{review_ESN}. Here, $\mathcal{P} ^{\mathrm{in} } _o $, $\mathcal{P} ^{\mathrm{in} } _a $, $\mathcal{P} ^{\mathrm{in} } _b $, and $\mathcal{P} ^{\mathrm{res} } _0 $, the distribution of nonzero components of $\vector{W} ^{\mathrm{in} } _o $, $\vector{W} ^{\mathrm{in} } _a $, $\vector{W} ^{\mathrm{in} } _b $, and $\vector{W} ^{\mathrm{res} } _0 $ were expressed as gaussian functions $ \mathcal{N} \left( 0, \left( \sigma^{\mathrm{in} } _o \right) ^2 \right)$, $\mathcal{N} \left( 0, \left( \sigma^{\mathrm{in} } _a \right) ^2 \right)$, $\mathcal{N} \left( 0, \left( \sigma^{\mathrm{in} } _b \right) ^2 \right)$, and $\mathcal{N} \left( 0, \left( \sigma^{\mathrm{res} } _0 \right) ^2 \right)$, respectively. The list of hyperparameters are presented in Table \ref{hyperparameters}.

\begin{table}
\centering
\begin{tabular}{ccc}
 character & meaning & value \\ \hline\hline
 $N^{\mathrm{res} }$ & the number of neurons of the reservoir & 1024 \\ \hline
 $\alpha$ & leaking rate & 0.8 \\ \hline
 $p_{s1} ^{\mathrm{in} }$ & sparsity of $\vector{W} ^{\mathrm{in} } _o$ corresponding to $\mathcal{A} _1$ & 0.6 \\ \hline
 $p_{s2} ^{\mathrm{in} }$ & sparsity of $\vector{W} ^{\mathrm{in} } _o$ corresponding to $\mathcal{A} _2 \backslash \mathcal{A} _1$ & 0.8 \\ \hline
 $p_{s3} ^{\mathrm{in} }$ & sparsity of $\vector{W} ^{\mathrm{in} } _o$ corresponding to $\mathcal{A} _3 \backslash \mathcal{A} _2$ & 0.9 \\ \hline
 $p_{sb} ^{\mathrm{in} }$ & sparsity of $\vector{W} ^{\mathrm{in} } _b$ & 0.9 \\ \hline
 $p_{s} ^{\mathrm{res} }$ & sparsity of $\vector{W} ^{\mathrm{res} }$ & 0.9 \\ \hline
 $\sigma ^{\mathrm{in} } _o $ & stantard deviation of $\mathcal{P} ^{\mathrm{in} } _o $ & 1.0 \\ \hline
 $\sigma ^{\mathrm{in} } _a $ & stantard deviation of $\mathcal{P} ^{\mathrm{in} } _a $ & 2.0 \\ \hline
 $\sigma ^{\mathrm{in} } _b $ & stantard deviation of $\mathcal{P} ^{\mathrm{in} } _b $ & 1.0 \\ \hline
 $\sigma ^{\mathrm{res} } _0 $ & stantard deviation of $\mathcal{P} ^{\mathrm{res} } _0 $ & 1.0 \\ \hline
 $\rho$ & spectral radius of $\vector{W} ^{\mathrm{res} }$ & 0.95 \\ \hline
 $t_{\mathrm{max} }$ & number of time steps during one episode & 500 \\ \hline
 $\gamma$ & discount factor & 0.95 \\ \hline
 $\epsilon _0$ & initial value of $\epsilon$ & 1.0 \\ \hline
 $\delta _{\epsilon}$ & decay rate of $\epsilon$ & 0.95 \\ \hline
 $\epsilon _{\mathrm{min} } $ & threshold value that stop decaying $\epsilon$ & 0.02 \\ \hline
 $\lambda$ & forgetting factor & 0.95 \\ \hline
 
 $\beta$ & coefficient used for the initial value of $\tilde{A}_G $ & $1 \times 10^{-4}$ \\ \hline
\end{tabular}
\vspace{1.0mm}
\caption{Hyperparameters of this study}
\label{hyperparameters}
\end{table}
As the specific tasks, we considered the following two situations, changing the number of agents, $n_{\mathrm{agent} }$. Note that we did not execute any kinds of pretraining in the simulations. 

\subsubsection{Task.~I : Choice between a narrow direct route and a broad detour \label{detour_setting} }
\begin{figure}[!tb]
 \centering
\includegraphics[width = 7.0cm]{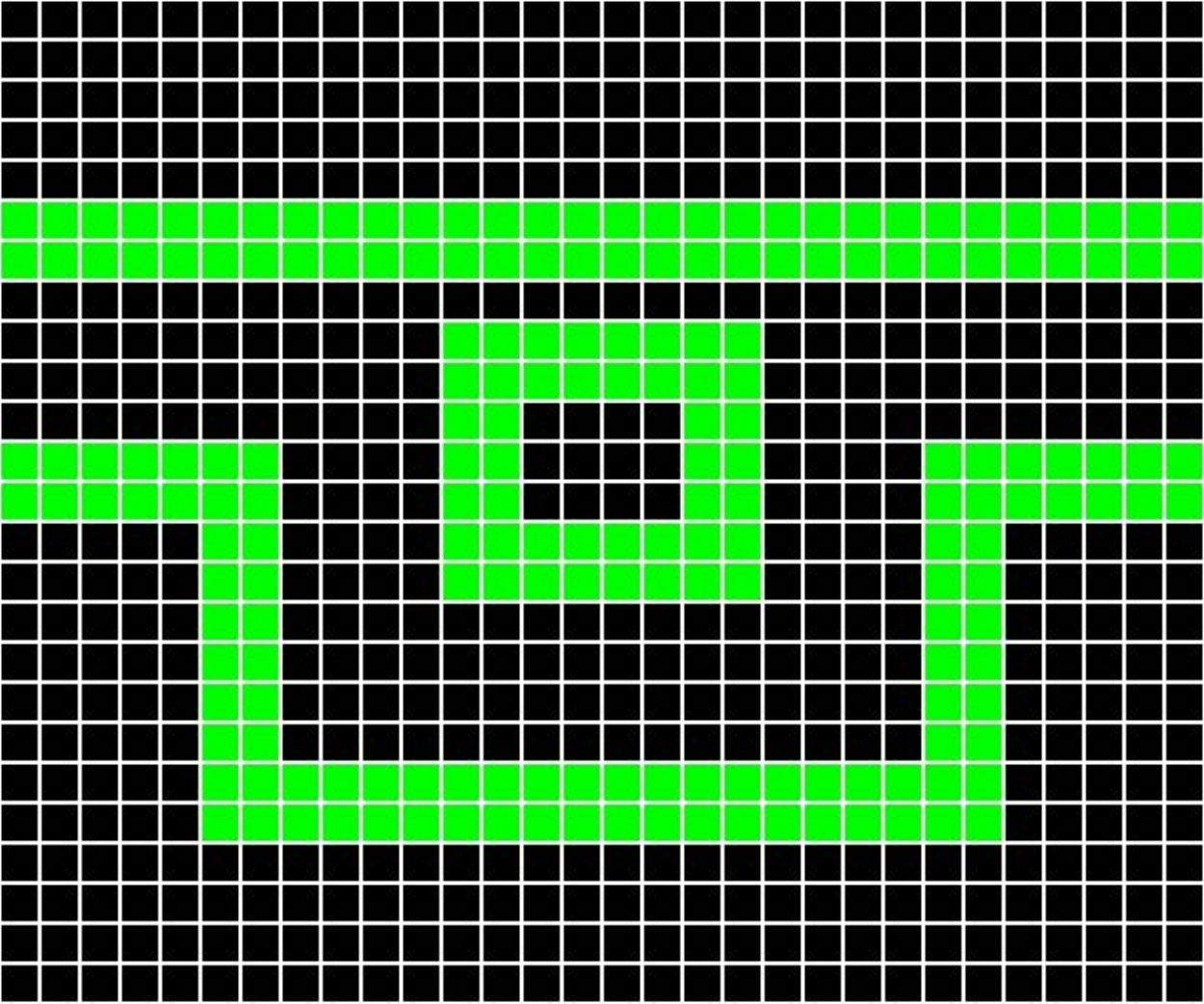}
\caption{Forked road considered in task.~I. Grids indicating vacant areas and walls are painted black and green, respectively. White lines are drawn to emphasize the borders of grids. }
\label{detour_grid}
\end{figure}

\begin{figure}[!tb]
 \centering
\includegraphics[width = 8.0cm]{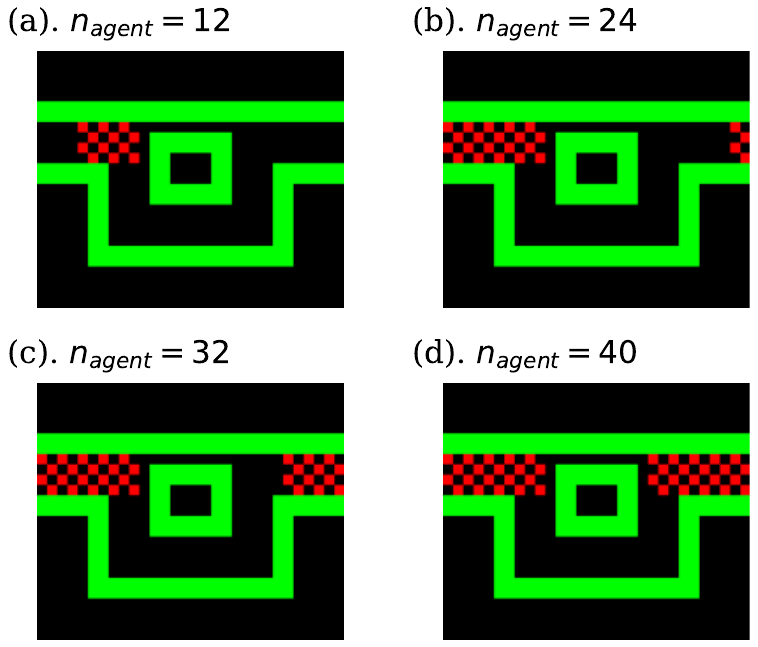}
\caption{Initial placement of agents at (a).$n_{\mathrm{agent}} = 12$, (b).$n_{\mathrm{agent}} = 24$, (c).$n_{\mathrm{agent}} = 32$ and (d)$n_{\mathrm{agent}} = 40$ in the task.~I. Meanings of the black and green cells are the same as Fig.~\ref{detour_grid}, and agents are painted red. }
\label{init_detour}
\end{figure}

We first considered a forked road (Fig.~\ref{detour_grid}), which was composed of a narrow direct route and a broad detour. The right and left edges were connected by the periodic boundary condition. In the initial state, agents were arranged in the checkerboard pattern on the part where the road was not forked, as shown in Fig.~\ref{init_detour}.

Here, the case wherein every agent attempted to go right was considered. Thus, we let the reward of $i$-th agent, $r_{i;t}$, be +1(-1) when it went right(left), and 0 otherwise. The number of groups was set as 1 because the aim of the agents was similar. Namely, all agents shared one policy, and all of their experiences were reflected in the update of $\tilde{A}_G $, $\tilde{B}_G $, and $\vector{W} _i ^{\mathrm{out} }$. The sharing of the experiences was realized only through Eqs.~(\ref{Wout_ESN}), (\ref{A_ESN}), and (\ref{B_ESN}), and the reward of each agent itself was not affected by those of other agents. In this environment, agents were believed to go through the direct route when they were not disturbed by other agents. However, if the number of agents increases and the road becomes crowded, limited number of agents can use the direct route, and others should choose the detour. We investigated whether the agents could learn this choice between two routes. 

\subsubsection{Task.~II : Bidirectional pedestrian flow in a corridor \label{lane_setting} }
\begin{figure}[!tb]
 \centering
\includegraphics[width = 8.0cm]{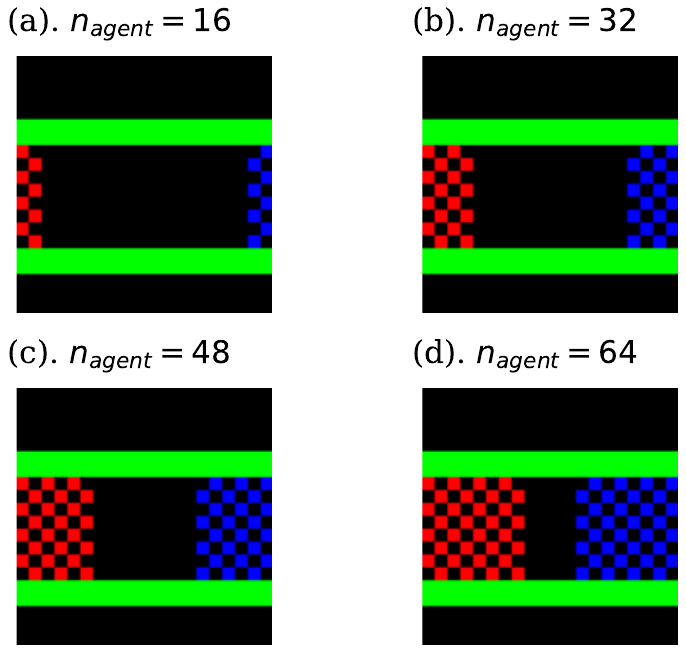}
\caption{Initial placement of agents at (a).$n_{\mathrm{agent}} = 16$, (b).$n_{\mathrm{agent}} = 32$, (c).$n_{\mathrm{agent}} = 48$ and (d)$n_{\mathrm{agent}} = 64$ in the task.~II. Meanings of the black and green cells are the same as Fig.~\ref{detour_grid}, and right(left)-proceeding agents are painted red(blue). 
In this figure, agents of different groups are painted in different colors so that we can distinguish them from each other. However, they have the same color, $(1,0)$, in the viewpoint of agents. }
\label{init_broadlane}
\end{figure}

We next considered the case where two types of agents, those trying to go right and left, existed in a corridor with width 8 and length 20. As in the task.~I, the right and left edges were connected by the periodic boundary condition, and the agents were arranged in the checkerboard pattern in the initial state, as shown in Fig.~\ref{init_broadlane}.
Here, we divided the agents into two groups, those of the right and left-proceeding agents. The reward of the right-proceeding agent was the same as that of the previous case. Whereas, that of the left-proceeding agent was -1(+1) when it went right(left), and 0 otherwise. Note that the parameter sharing through Eq. (\ref{Wout_ESN}) was executed only among the agents of the same group.
In this case, agents should go through the corridor avoiding oppositely-proceeding agents. We investigated whether and how agents achieve this task.

\section{Results \label{results}}
In this section, we present the results of the simulations explained in the previous section.

\subsection{Performance in task.~I \label{result_detour} }
\begin{figure}[!tb]
 \centering
\includegraphics[width = 12.0cm]{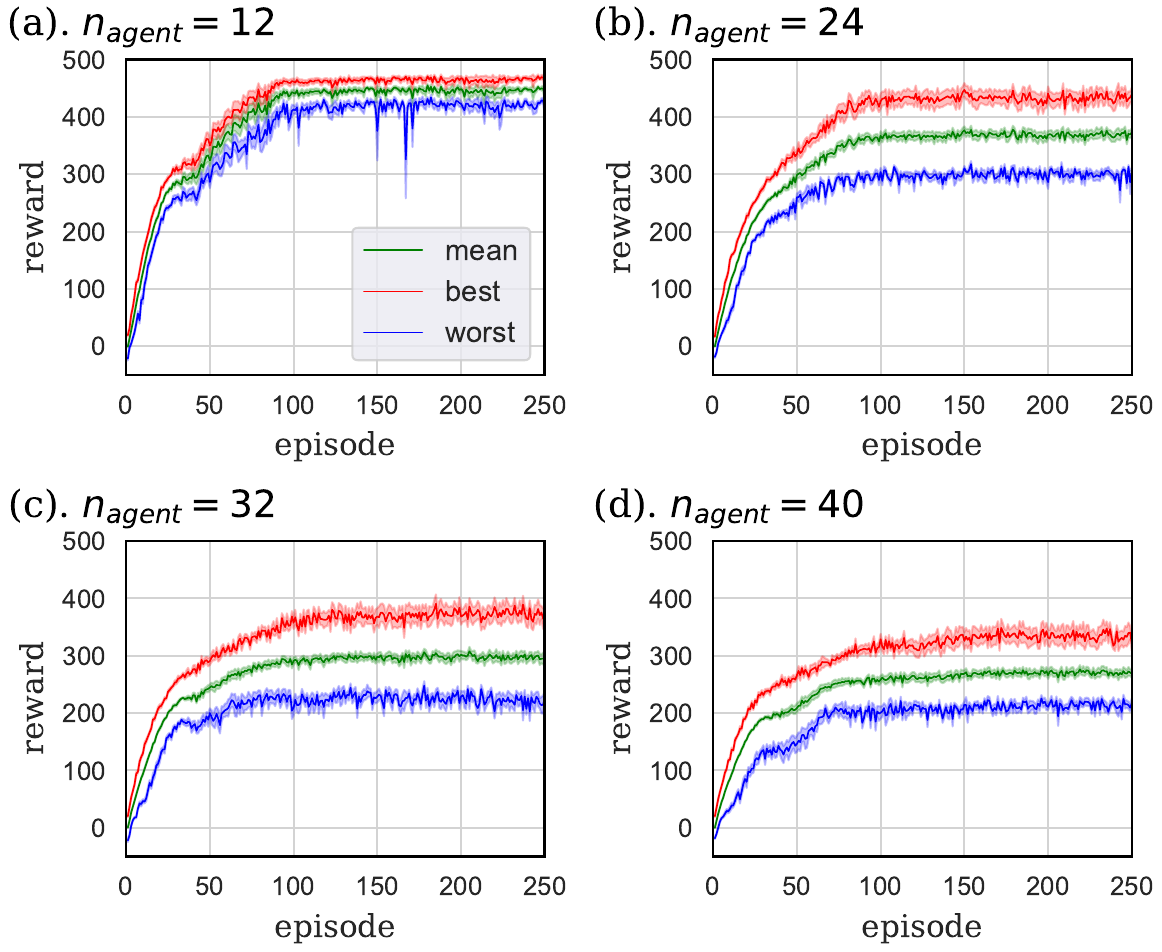}
\caption{Learning curves of task.~I at (a).$n_{\mathrm{agent}} = 12$, (b).$n_{\mathrm{agent}} = 24$, (c).$n_{\mathrm{agent}} = 32$ and (d)$n_{\mathrm{agent}} = 40$. The green curve is the mean value of all agents' rewards, and the red and blue curves indicate the rewards of the best and worst-performing agents. Each value is averaged over 8 independent trials, and the standard errors taken from them are painted in pale colors. }
\label{LC_detour}
\end{figure}

\begin{figure}[!tb]
 \centering
\includegraphics[width = 8.0cm]{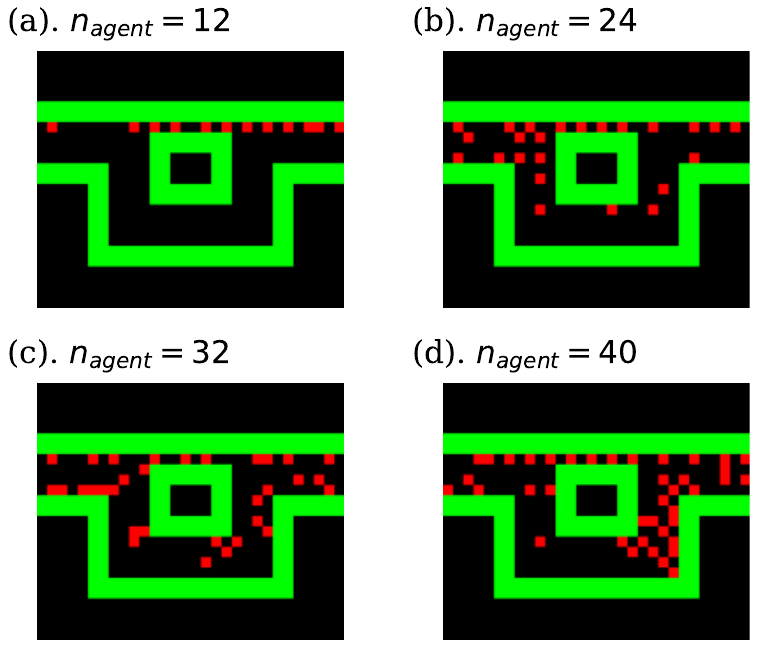}
\caption{Snapshots of task.~I at $t=100$ in the case that (a).$n_{\mathrm{agent}} = 12$, (b).$n_{\mathrm{agent}} = 24$, (c).$n_{\mathrm{agent}} = 32$, and (d).$n_{\mathrm{agent}} = 40$. In these figures, agents have finished 250 episodes training (i.e. these figures are taken at the 251th episode.) Meanings of the cells are the same as Fig.~\ref{init_detour}. }
\label{snap_detour}
\end{figure}

\begin{figure}[!tb]
 \centering
\includegraphics[width = 10.0cm]{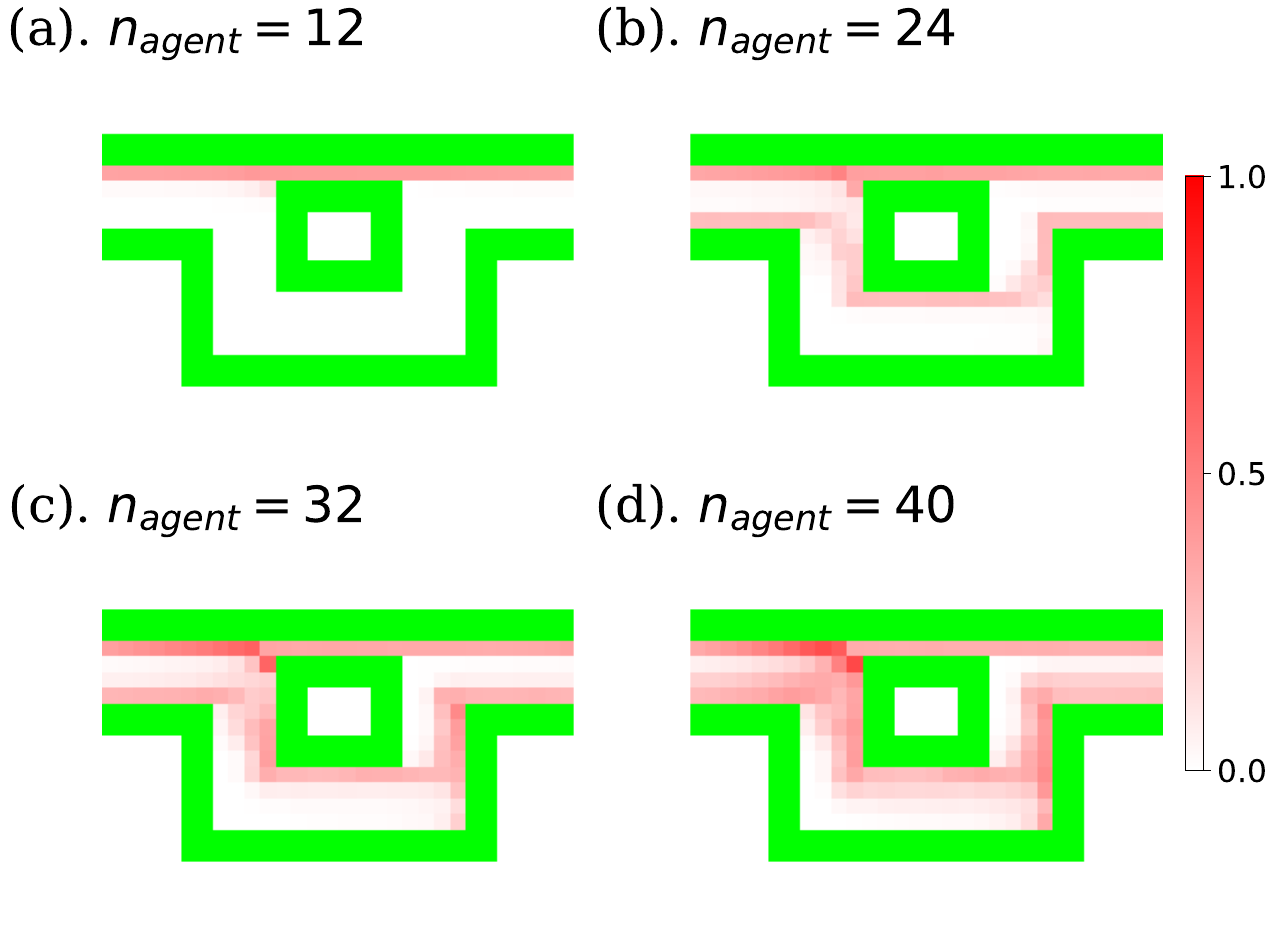}
\caption{Density colormaps of task.~I averaged over $100 \leq t \leq t_{\mathrm{max} }-1(=499)$ of the 151–250 episodes of 8 independent trials in the case that (a).$n_{\mathrm{agent}} = 12$, (b).$n_{\mathrm{agent}} = 24$, (c).$n_{\mathrm{agent}} = 32$, and (d).$n_{\mathrm{agent}} = 40$.  }
\label{colormap_detour}
\end{figure}

The learning curves of task.~I are shown in Fig.~\ref{LC_detour}. In these graphs, the green curve is the mean value of all agents' rewards, and the red and blue curves indicate the rewards of the best and worst-performing agents. In each case, we executed 8 independent trials and averaged the values of the graphs, and painted the standard errors taken from these trials in pale colors. As shown in Fig.~\ref{LC_detour}, the performance of each agent appeared to worsen with increasing $n_{\mathrm{agent} }$. This is because each agent was prevented from proceeding by other agents. In the case that $n_{\mathrm{agent} } \geq 24$, the direct route was excessively crowded and certain agents had no choice but to go to the detour, as shown in snapshots and density colormaps in Figs.~\ref{snap_detour} and \ref{colormap_detour}.~(b)-(d). Thus, the difference in the performance between the best and worst agents increased compared to the case where $n_{\mathrm{agent} } =12$. In addition, seeing the density colormaps of Fig.~\ref{colormap_detour}, some agents tend to get stuck in the bottom-right corner of the detour when $n_{\mathrm{agent} }$ is large. Considering that there is no benefit for agents in staying at this corner, it is thought to be the misjudgment resulting from the low information processing capacity of ESN.

\begin{figure}[!tb]
\centering
\includegraphics[width = 5.0cm]{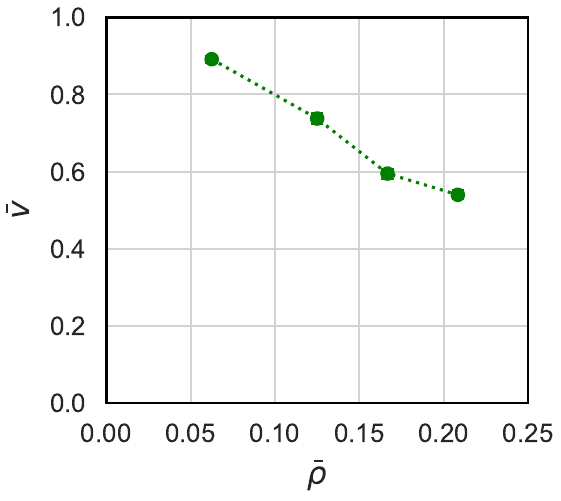}
\caption{Fundamental diagram of task.~I averaged over the 151–250 episodes of 8 independent trials. Dotted lines are guides to the eyes. }
\label{fund_graph_detour}
\end{figure}

From the definition of reward in this study, the total reward that one agent gets during one episode is equal to the displacement along the direction it tries to go. Hence, the average velocity of agents, $\bar{v}$, can be calculated as
\begin{equation}
\bar{v} = \frac{\left( \mathrm{mean \ value \ of \ all \ agents' \ rewards  } \right) }{t_{\mathrm{max} } } . \label{bar_v}
\end{equation}
Note that the upper bound of this value is 1, which can be realized when all agents always proceed in the direction they want to. In addition, considering that simulations of this study impose the periodic boundary condition and the number of agents $n_{\mathrm{agent}} $ does not change, the average density $\bar{\rho}$ is expressed as
\begin{equation}
\bar{\rho} = \frac{n_{\mathrm{agent}}  }{\left( \mathrm{the \ number \ of \ cells \ that \ agents \ can \ walk \ into  } \right) } . \label{bar_rho}
\end{equation}
Here, note that the phrase ``the number of cells that agents can walk into'' means the area that is not devided from agents by walls, and does not mean the number of cells that a certain agent can move into at a certain step. In the case of task.~I, for example, this number can be counted as 192, by Fig.~\ref{detour_grid}.
Using these equations, we can plot the fundamental diagram\cite{SSKB05}, the graph that draw the relation between velocity $\bar{v}$ and density $\bar{\rho}$, as Fig.~\ref{fund_graph_detour}. These data are averaged over the 151-250 episodes of 8 independent trials. In this graph, we estimated the error bars by calculating the standard error of the mean over 8 independent trials, but these error bars are smaller than the points. Seeing Fig.~\ref{fund_graph_detour}, $\bar{v}$ has the value near its upper bound when $\bar{\rho}$ is small, and gradually decreases with increasing $\bar{\rho}$. This decrease itself is observed in many situations of pedestrian dynamics. However, considering that  agents get stuck in the corner under present calculation, the improvement of the algorithm and neural network is thought to be required before quantitative comparisons with previous studies.

\subsection{Performance in task.~II \label{result_lane} }
\begin{figure}[!tb]
 \centering
\includegraphics[width = 12.0cm]{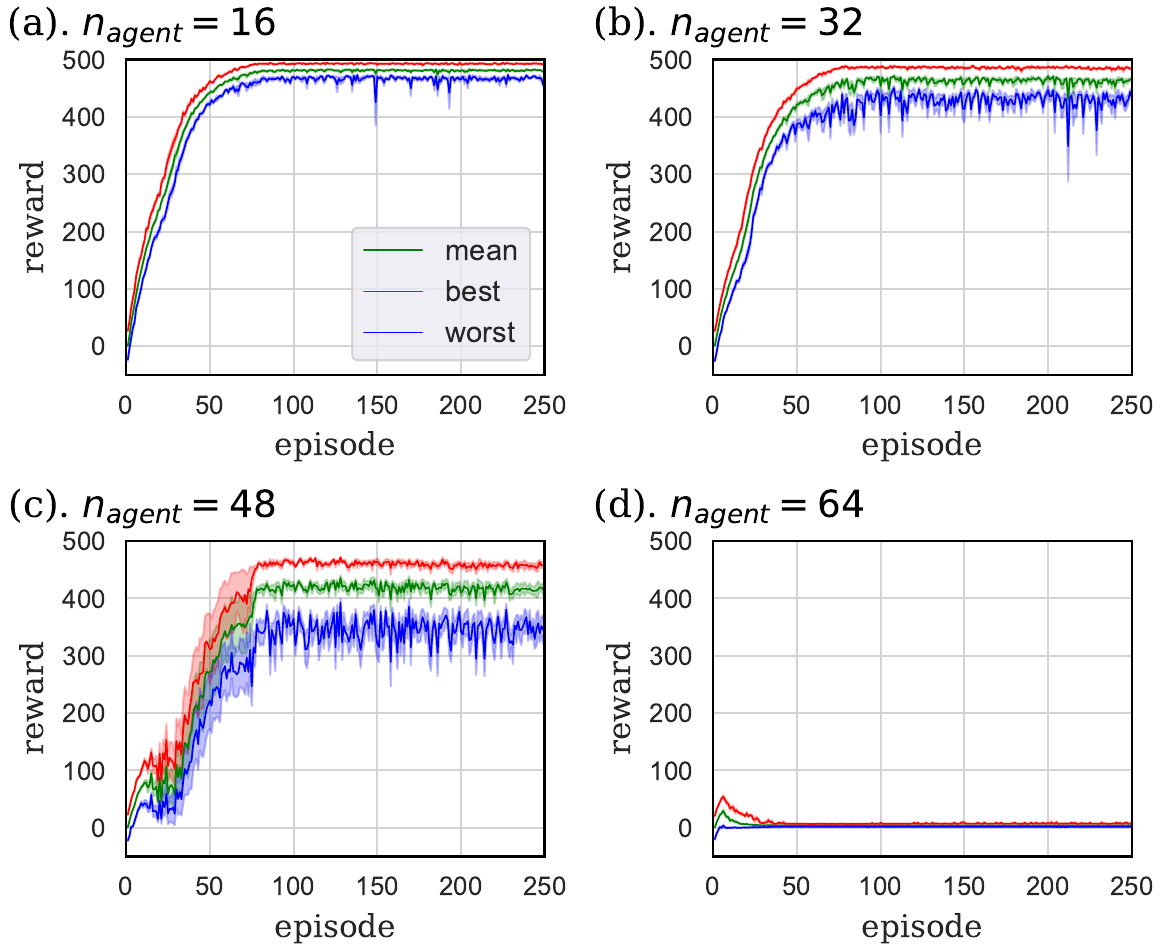}
\caption{Learning curves of task.~II at (a).$n_{\mathrm{agent}} = 16$, (b)$n_{\mathrm{agent}} = 32$, (c).$n_{\mathrm{agent}} = 48$, and (d). $N_{\mathrm{agent}} = 64$. Meaning of curves are the same as Fig.~\ref{LC_detour}. Each value is averaged over 8 independent trials, and the standard errors taken from them are painted in pale colors. }
\label{LC_lane}
\end{figure}

\begin{figure}[!tb]
 \centering
\includegraphics[width = 8.0cm]{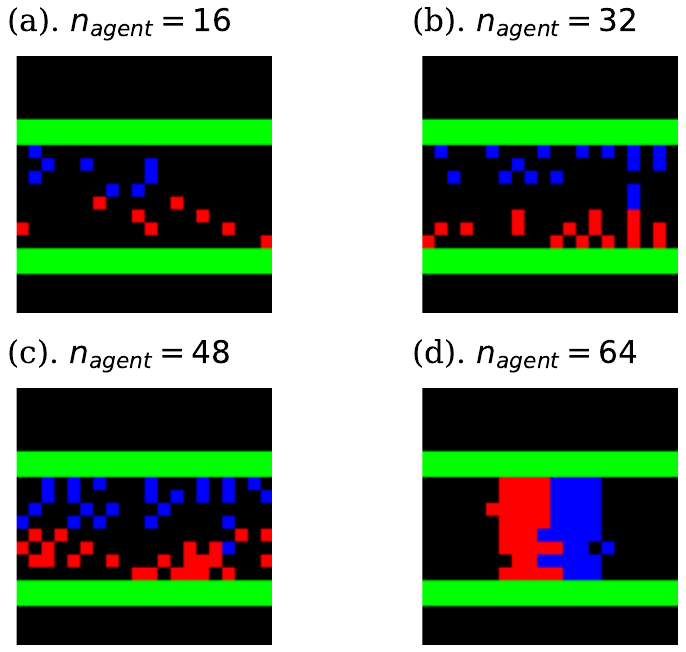}
\caption{Snapshots of task.~II at $t=100$ in the case that (a).$n_{\mathrm{agent}} = 16$, (b).$n_{\mathrm{agent}} = 32$, (c).$n_{\mathrm{agent}} = 48$, and (d).$n_{\mathrm{agent}} = 64$. In these figures, agents have finished 250 episodes training. Meanings of the cells are the same as Fig.~\ref{init_broadlane}. }
\label{snap_broadlane}
\end{figure}

\begin{figure}[!tb]
 \centering
\includegraphics[width = 5.5cm]{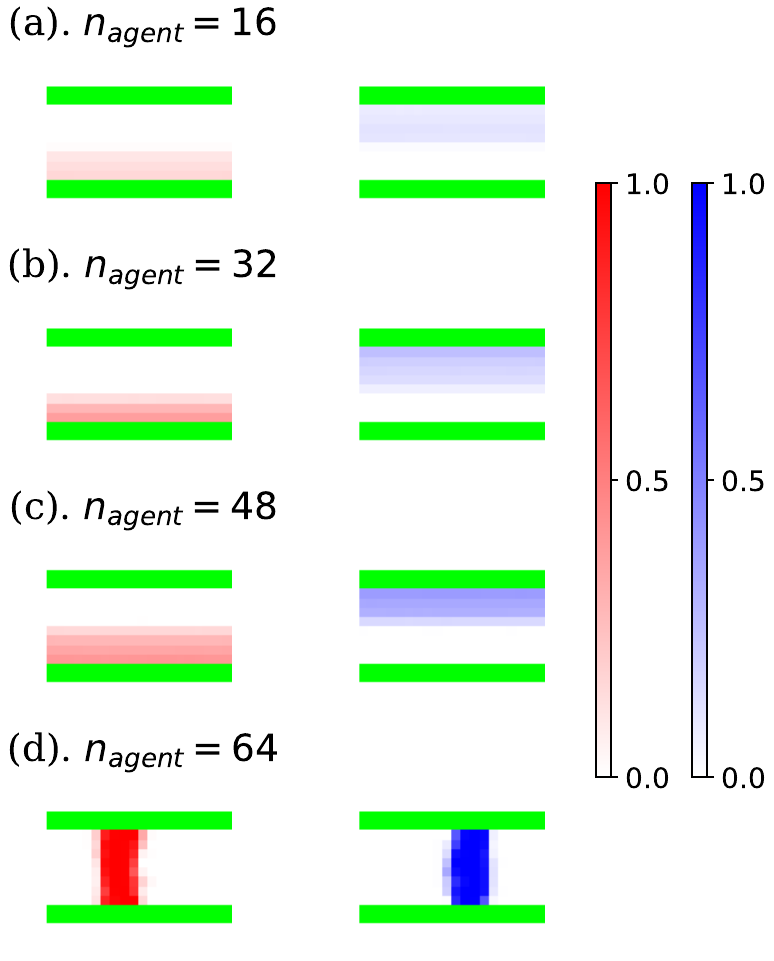}
\caption{Density colormaps of task.~II averaged over $100 \leq t \leq t_{\mathrm{max} }-1(=499)$ of the 151–250 episodes in the case that (a).$n_{\mathrm{agent}} = 16$, (b).$n_{\mathrm{agent}} = 32$, (c).$n_{\mathrm{agent}} = 48$, and (d).$n_{\mathrm{agent}} = 64$.  The left(right) columns indicate the data of right(left)-proceeding agents. These figures are taken from 1 trial because the lanes have different shapes between trials with different random seeds. }
\label{colormap_broadlane}
\end{figure}

\begin{figure}[!tb]
 \centering
\includegraphics[width = 8.0cm]{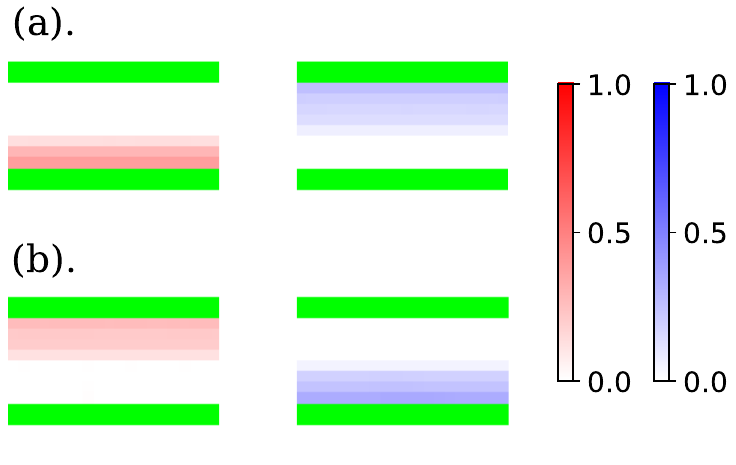}
\caption{Examples of density colormaps of task.~II in the case that $n_{\mathrm{agent}} = 32$ with different random seeds, drawn by the same way as Fig.~\ref{colormap_broadlane}. }
\label{colormap_broadlane32}
\end{figure}

The learning curves of task.~II are shown in Fig.~\ref{LC_lane}. Here, as discussed in Section \ref{result_detour}, we averaged the values of the rewards over 8 independent trials, and painted the standard errors among these trials in pale colors. According to Fig.~\ref{LC_lane}, the training itself was successful except when $n_{\mathrm{agent} } = 64$. The corresponding snapshots and density colormaps shown in Figs.~\ref{snap_broadlane} and \ref{colormap_broadlane} show that in the case of $n_{\mathrm{agent} } \leq 48$, agents of the same group created lanes and avoided collisions with agents in the opposite direction. Such lane formation was also observed in various previous studies that used mathematical models\cite{HM95}, experiments\cite{FN16,KGKMS06,MFNN21}, and other types of RL agents \cite{MgLF14,HLXFLY23}. Note that the lanes had different shapes depending on the seed of random values, i.e. the right-proceeding agents walked on the upper side of the corridor in some trials, and on the lower side in others, for example. Hence, the colormap of Fig.~\ref{colormap_broadlane} indicates the time average of one representative trial. To show the dependency on the random seed, we also drew the colormaps of the case that $n_{\mathrm{agent} } = 32$ with different seeds as Fig.~\ref{colormap_broadlane32}. 

When $n_{\mathrm{agent} } = 64$, two groups of agents failed to learn how to avoid each other and collided. In this case, each agent cannot move forward after the collision, and the reward they obtained was limited to that for the first few steps. 
Note that in this task, each agent cannot access the experiences of agents of the other group. Thus, thinking from the perspective of MARL, there is a possibility that the non-stationarity of the environment (in the viewpoint of agents) worsened the learning efficiency, as with the case of the independent learners.
However, from the perspective of soft matter physics, this stagnation of the movement under high density resembles the jamming transition observed in wide range of granular or active matters\cite{MIN99,oHSLN03,MSLB07,HFM11}. Nevertheless, to investigate whether it is really related with the jamming transition, simulations with larger scale are required. 

\begin{figure}[!tb]
\centering
\includegraphics[width = 5.5cm]{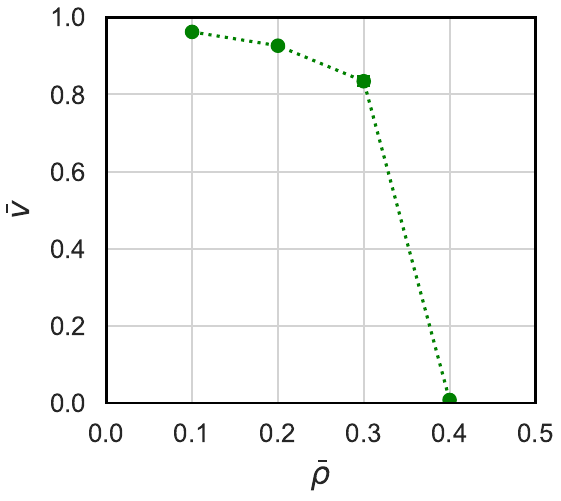}
\caption{Fundamental diagram of task.~I averaged over the 151–250 episodes of 8 independent trials. Dotted lines are guides to the eyes. }
\label{fund_graph_broadlane}
\end{figure}

As in task.~I, we can plot the fundamental diagram shown in Fig.~\ref{fund_graph_broadlane} using Eqs.~(\ref{bar_v}) and (\ref{bar_rho}). This graph shows the sudden decrease of the average velocity between $n_{\mathrm{agent}} = 48$ and 64 (i.e. $\bar{\rho} = 0.3$ and 0.4), reflecting the stagnation of the movement discussed above. Similar behavior is also reported in previous researches that considered the jamming transition in mathematical models of pedestrians\cite{MIN99}.

\subsection{Comparison with independent learners \label{independent} }
As explained in Section \ref{ESN_LSPI}, we adopted the parameter sharing method in our calculations. In this section, we calculate the two tasks discussed above considering the case that the parameter sharing was not adopted, and compare the results with those of previous sections to investigate the effect of this method. Specifically, equations
\begin{equation}
\vector{W} _i ^{\mathrm{out} } = \tilde{B}_{i} \tilde{A} _{i} ^{-1} , \label{Wout_ESN_indep}
\end{equation} 
\begin{eqnarray}
\tilde{A} _{i} & = & \sum _{n=1} ^{n_l} \lambda^{n_{l} -n} \sum _{t \in \mathcal{E}_n } \left( \hat{ \vector{X} } _i ^{\mathrm{res} } (t) - \gamma \hat{ \vector{X} } _i ^{\mathrm{res} } (t+1) \right) \hat{ \vector{X} } _i ^{\mathrm{res} } (t) ^T \nonumber \\
& & + \lambda ^{n_l - 1} \beta I , \label{A_ESN_indep} 
\end{eqnarray}
and
\begin{equation}
\tilde{B} _{i} = \sum _{n=1} ^{n_l} \lambda^{n_{l} -n} \sum _{t \in \mathcal{E}_n } r_{i;t} \hat{ \vector{X} } _i ^{\mathrm{res} } (t) ^T , \label{B_ESN_indep} 
\end{equation}
were used to calculate $\vector{W} _i ^{\mathrm{out} }$ in this section. Thus, each agent learned independently. To observe the effect of replacing Eq.(\ref{Wout_ESN}) by Eq.(\ref{Wout_ESN_indep}), input and reservoir weight matrices, $\vector{W} ^{\mathrm{in} } _o $, $\vector{W} ^{\mathrm{in} } _a $, $\vector{W} ^{\mathrm{in} } _b $, and $\vector{W} ^{\mathrm{res} } $, were still shared by all agents, and the hyperparameters were the same as listed in Table \ref{hyperparameters}.

\begin{figure}[!tb]
 \centering
\includegraphics[width = 12.0cm]{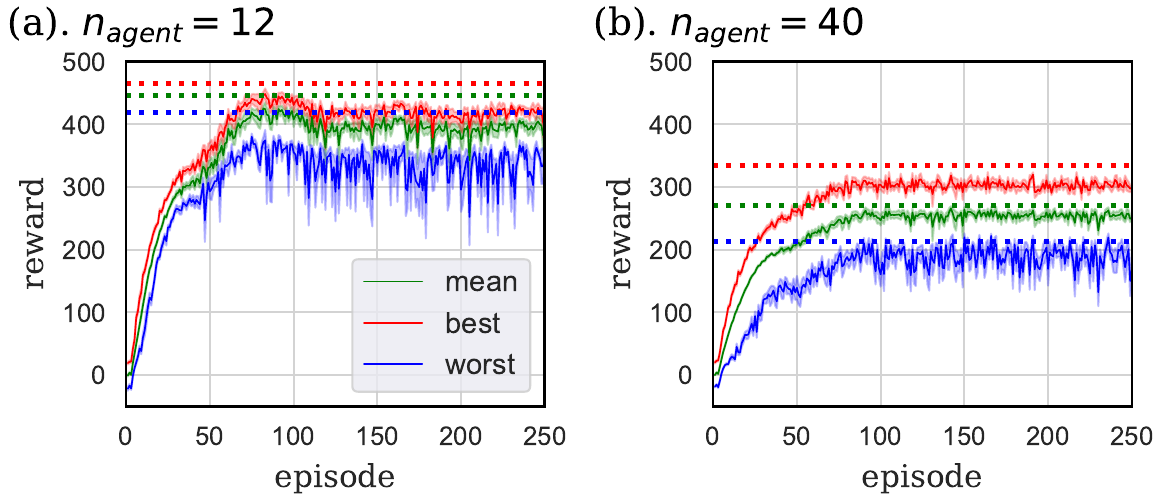}
\caption{Learning curves of independent learners in task.~I at (a).$n_{\mathrm{agent}} = 12$ and (b).$n_{\mathrm{agent}} = 40$. Meaning of curves are the same as Fig.~\ref{LC_detour}, and the dotted lines are the corresponding scores under the parameter sharing method averaged over the 151–250 episodes of 8 independent trials.}
\label{LC_detour_indep}
\end{figure}

\begin{figure}[!tb]
 \centering
\includegraphics[width = 12.0cm]{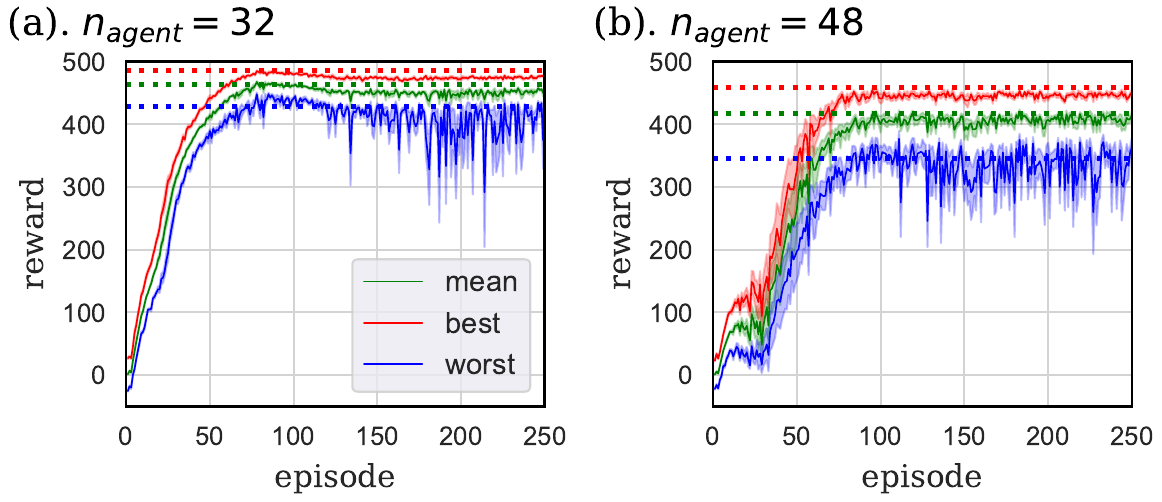}
\caption{Learning curves of independent learners in task.~II at (a).$n_{\mathrm{agent}} = 32$ and (b).$n_{\mathrm{agent}} = 48$. Meaning of curves and dotted lines are the same as Fig.~\ref{LC_detour_indep}.}
\label{LC_lane_indep}
\end{figure}
The learning curves of tasks.~I and II are shown in Figs.~\ref{LC_detour_indep} and \ref{LC_lane_indep}, respectively. In these graphs, scores when using the parameter sharing method averaged over the 151–250 episodes of 8 independent trials are drawn as the dotted lines. A comparison of the learning curves with these dotted lines revealed that the performance when the parameter sharing was not used was almost the same as or slightly lower than that under the parameter sharing. Hence, the parameter sharing method did not result in the drastic improvement of the learning. The advantage of this method is rather the reduction in the number of inverse matrix calculations, at least in the case of our tasks. We actually calculated the case of task.~II under $n_{\mathrm{agent}} = 64$, but did not plot the graph because agents failed to make lanes like the case of Section \ref{result_lane}.

\subsection{Comparison with the case that two groups share the parameters in task.~II \label{1group} }
As we explained in Section \ref{lane_setting}, in the case of task.~II, we divided the pedestrians into two groups by the direction they want to proceed in. Here, the training of each group was executed independently. However, we can also implement the MARL agents even in the case that the parameters of neural networks are shared between two groups. Specifically, we first define a two-dimensional one-hot vector $\vector{\eta} _g = (1,0)^T$ or $(0,1)^T$, which represents the information on each agent's group. 
Adding this vector to agents' observation and modifying eq.~(\ref{tildeXres_proposed}) as follows:
\begin{equation}
\tilde{ \vector{X} } _i ^{\mathrm{res} } (t;a) = f \left( \vector{W} ^{\mathrm{in} } _o \vector{o}_{i;t}   + \vector{W} ^{\mathrm{in} } _a \vector{\eta} _a + \vector{W} ^{\mathrm{in} } _g \vector{\eta} _g + \vector{W} ^{\mathrm{in} } _b + \vector{W} ^{\mathrm{res} } \vector{X} _i ^{\mathrm{res} } (t-1) \right) , \label{tildeXres_proposed_1group} 
\end{equation}
the learning process can be executed even if parameters are shared between two groups. To share the parameters, equations
\begin{equation}
\vector{W} _i ^{\mathrm{out} } = \tilde{B} \tilde{A} ^{-1} , \label{Wout_ESN_1group}
\end{equation}
\begin{eqnarray}
\tilde{A} & = & \sum _{n=1} ^{n_l} \lambda^{n_{l} -n} \sum _{j : \mathrm{all \ agents} } \sum _{t \in \mathcal{E}_n }  \left( \hat{ \vector{X} } _j ^{\mathrm{res} } (t) - \gamma \hat{ \vector{X} } _j ^{\mathrm{res} } (t+1) \right) \hat{ \vector{X} } _j ^{\mathrm{res} } (t) ^T \nonumber \\
& & + \lambda ^{n_l - 1} \beta I \label{A_ESN_1group} ,
\end{eqnarray}
and
\begin{equation}
\tilde{B} = \sum _{n=1} ^{n_l} \lambda^{n_{l} -n} \sum _{j : \mathrm{all \ agents} } \sum _{t \in \mathcal{E}_n } r_{j;t} \hat{ \vector{X} } _j ^{\mathrm{res} } (t) ^T , \label{B_ESN_1group} 
\end{equation}
are used to calculate $\vector{W} _i ^{\mathrm{out} }$, instead of eqs.~(\ref{Wout_ESN}), (\ref{A_ESN}), and (\ref{B_ESN}).
 Here, we let $\vector{W} ^{\mathrm{in} } _g$, the input weight matrix corresponding to $\vector{\eta} _g$, be a dense matrix, and used the same standard deviation as $\vector{W} ^{\mathrm{in} } _a$. Namely, the distribution of each component of $\vector{W} ^{\mathrm{in} } _g$ is $\mathcal{N} \left( 0, \left( \sigma^{\mathrm{in} } _a \right) ^2 \right)$. In this section, we investigate whether this method work. The hyperparameters themselves were the same as listed in Table \ref{hyperparameters}.

\begin{figure}[!tb]
 \centering
\includegraphics[width = 12.0cm]{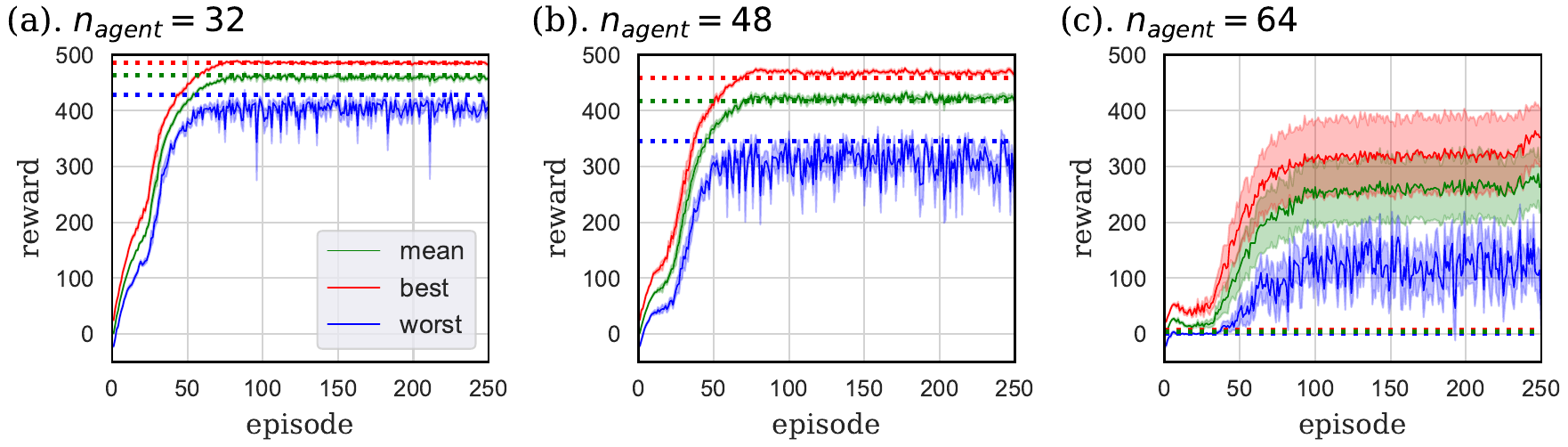}
\caption{Learning curves of the case that two groups share the parameters of the neural network in task.~II at (a).$n_{\mathrm{agent}} = 32$, (b).$n_{\mathrm{agent}} = 48$, and (c).$n_{\mathrm{agent}} = 64$. Meaning of curves is the same as Fig.~\ref{LC_detour_indep}, and the dotted lines are the corresponding scores of the case that two groups do not share their parameters averaged over the 151–250 episodes of 8 independent trials. }
\label{LC_lane_1group}
\end{figure}

The result is shown in Fig.~\ref{LC_lane_1group}. In this figure, the dotted lines are the scores of Section \ref{result_lane}, the case that two groups do not share their parameters, averaged over the 151–250 episodes of 8 independent trials. Seeing Fig.~\ref{LC_lane_1group} (a) and (b), in the case that $n_{\mathrm{agent}} = 32$ and 48, the parameter sharing between two groups slightly improved or hardly changed the mean and best scores compared with those under the training without parameter sharing between different groups, whereas the worst score was lowered. It is thought that the additional information, $\vector{\eta} _g$, led to the confusion of the neural network and wrong action choice, especially under subtle situations that is difficult to evaluate $Q_i (\vector{o}_{i;t},a)$. 
The result under $n_{\mathrm{agent}} = 64$ shown in Fig.~\ref{LC_lane_1group} (c) is interesting. As we saw in Section \ref{result_lane}, agents that did not share their parameters between different groups failed to learn to make lanes in all trials in this case. However, agents of this section succeeded in making lanes in some trials. We can think two possible factors that cause this difference. One is the number of agents choosing wrong actions. Blockage of agents in this task is caused when they learn wrong strategy that ``go straight and earn rewards for first few steps'', before they learn to make lanes. As we discussed above, agents of this section are thought to be sometimes confused by the additional information, $\vector{\eta} _g$. Even if agents learn the wrong strategy, blockage is eased compared with that of Section \ref{result_lane} because such ``confused'' agents do not simply go straight. The other possible factor is the sharing of the experience. Agents of this section share the experience between different groups by the parameter sharing, whereas those of Section \ref{result_lane} share it only among the same group. This experience sharing is thought to promote the efficient learning of correct strategy that ``agents of different groups should go to different lanes''. It is difficult to identify which of these two factors plays the significant role, only by this task.

\subsection{Comparison with deep reinforcement learning (DRL) algorithms \label{DRL} }
In this section, we calculated the same tasks using representative DRL algorithms, DQN\cite{Mnih15}, advantage actor-critic(A2C)\cite{Mnih16}, and proximal policy optimization (PPO)\cite{Schulman17} methods, and compare the performance with our algorithm. In these calculations, we prepared the neural network with the same size as the reservoir of our ESN, i.e., the network has 1 hidden layer (of perceptron) with 1024 neurons. Each agent observed 2-channel bitmap data with $11 \times 11$ cells centering on itself and agents of the same group shared the experience by parameter sharing, as in our algorithm. Parameter sharing between different groups like Section \ref{1group} is not introduced. Note that in the A2C and PPO methods, actor- and critic-networks shared the parameters except for the output weight matrices. 

As the optimizer, we adopted Adam\cite{KB14}. The codes were implemented using Pytorch (torch 2.2.0), and we used the default values of this library for the initial distribution of the parameters of neural network and the hyperparameters of the optimizer except for the learning rate. The parameters related to environment or reward, such as $t_{\mathrm{max}}$ and $\gamma$, were the same as our method. In addition, we used the same value of $\epsilon$ as our method for DQN.
Other hyperparameters used for this section are listed in Table~\ref{hyperparameters_DRL}. Here, the learning rate for PPO was set smaller than those of other methods except the cases of task~II. with $n_{\mathrm{agent}} = 48$ and 64 to stabilize the learning. In addition, we let the minibatch size of DQN for task~II. with $n_{\mathrm{agent}} = 48$ and 64 larger than other tasks. The reason why we let these tasks be the exception is explained later.
\begin{table}
\centering
\begin{tabular}{cc}
meaning & value \\ \hline\hline
learning rate for DQN & $1 \times 10^{-3}$ for task~II. with $n_{\mathrm{agent}} = 48, 64$ \\ & $2.5 \times 10^{-4}$ otherwise \\ \hline
learning rate for A2C & $2.5 \times 10^{-4}$ \\ \hline
learning rate for PPO & $2.5 \times 10^{-4}$ for task~II. with $n_{\mathrm{agent}} = 48$ \\ 
  & $1 \times 10^{-3}$ for task~II. with $n_{\mathrm{agent}} = 64$ \\
  & $5 \times 10^{-5}$ otherwise \\ \hline
replay-memory size for DQN & $10^6$ \\ \hline
minibatch size for DQN & 1024 for task~II. with $n_{\mathrm{agent}} = 48, 64$ \\ 
 & 64 otherwise \\ \hline
number of steps between updates of network for A2C & 5 \\ \hline
number of steps between updates of network for PPO & 125 \\ \hline
replay-memory size for PPO & 500$n_{\mathrm{agent}}$ \\ \hline
minibatch size for PPO & 125 \\ \hline
number of epochs for PPO & 4 \\ \hline

value-loss coefficient for A2C and PPO & 0.5 \\ \hline
entropy coefficient for A2C and PPO & 0.01 \\ \hline
clipping value for PPO & 0.2 \\ \hline
$\lambda_{GAE}$ for PPO & 0.95 \\ \hline

maximum gradient norm for DQN and A2C & 50 \\ \hline
maximum gradient norm for PPO & 1 \\ \hline

 \end{tabular}
\vspace{1.0mm}
\caption{Hyperparameters for DRL methods}
\label{hyperparameters_DRL}
\end{table}

\begin{figure}[!tb]
 \centering
\includegraphics[width = 12.0cm]{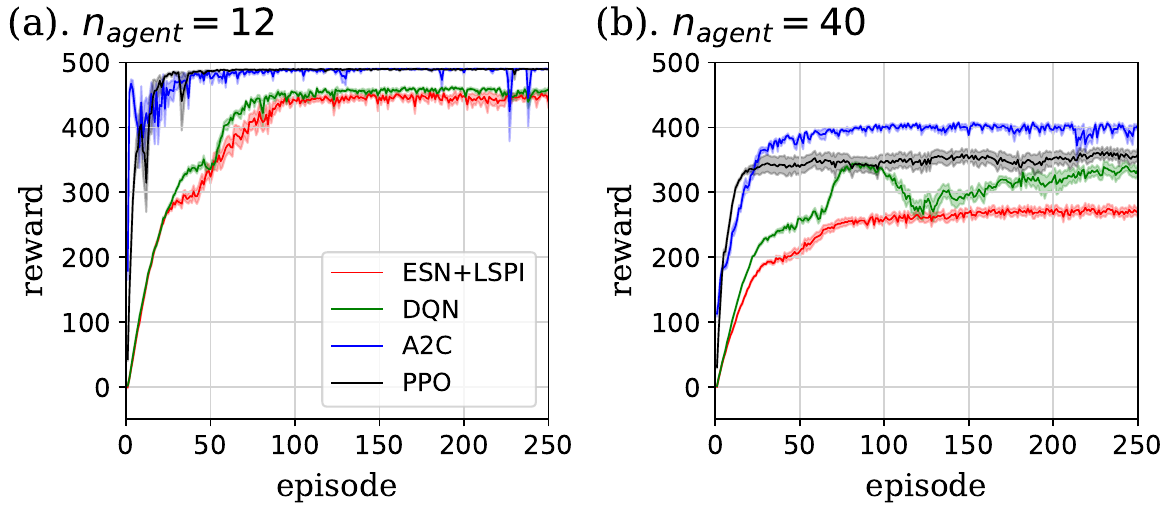}
\caption{Comparison of learning curves in task.~I at (a).$n_{\mathrm{agent}} = 12$ and (b).$n_{\mathrm{agent}} = 40$. The red, green, blue and black curves are the mean value of all agents' rewards of our method, DQN, A2C, and PPO, respectively. }
\label{LC_detour_DRL}
\end{figure}

\begin{figure}[!tb]
 \centering
\includegraphics[width = 12.0cm]{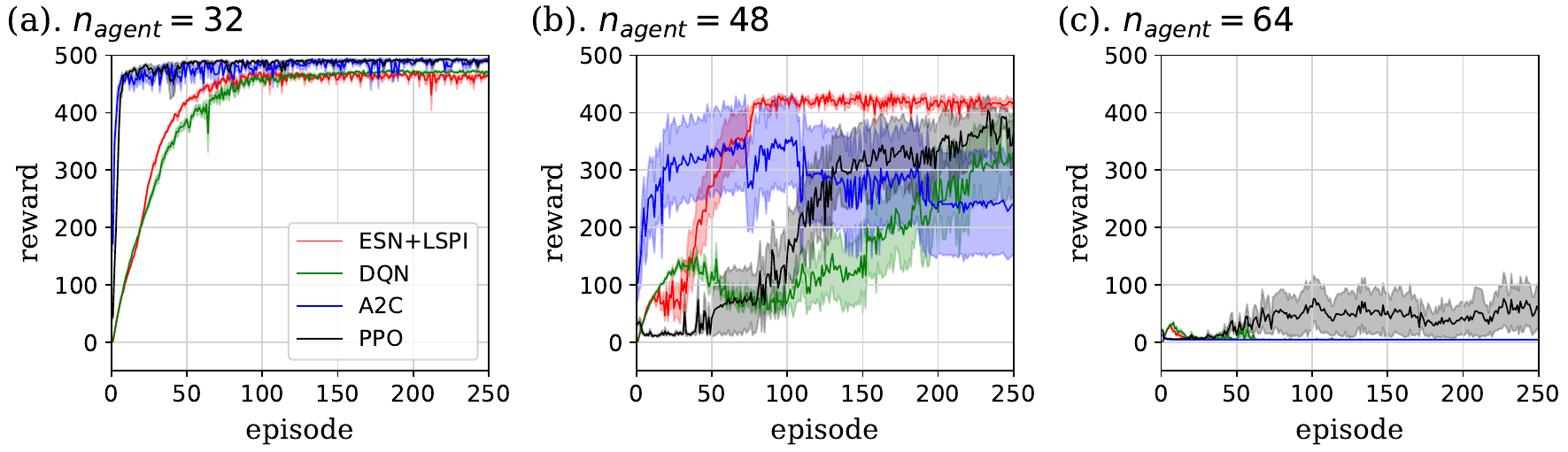}
\caption{Comparison of learning in task.~II at (a).$n_{\mathrm{agent}} = 32$, (b).$n_{\mathrm{agent}} = 48$ and (c).$n_{\mathrm{agent}} = 64$. Meaning of curves are the same as Fig.~\ref{LC_detour_DRL}.}
\label{LC_lane_DRL}
\end{figure}

The result is shown in Figs.~\ref{LC_detour_DRL} and \ref{LC_lane_DRL}. Here we plotted the mean value of all agents' rewards for each algorithm, and all graphs are averaged over 8 independent trials.
Seeing this figures, performance of our method on easy tasks, i.e. task.~I with $n_{\mathrm{agent}} = 12$ and task.~II with $n_{\mathrm{agent}} = 32$ resembles that of DQN, and slightly worse than the policy gradient methods. In addition, the learning speed of ours is slower than these methods. These points are thought to result from the $\epsilon$-greedy method, which decreases $\epsilon$ gradually and keeps this constant larger than the threshold value, $\epsilon _{\mathrm{min} }$. However, the performance on the task.~I with $n_{\mathrm{agent}} = 40$ is explicitly worse than all DRL algorithms. As we discussed in Sec.~\ref{result_detour}, the low information processing capacity of ESN is thought to be the cause of this phenomenon.

\begin{figure}[!tb]
 \centering
\includegraphics[width = 12.0cm]{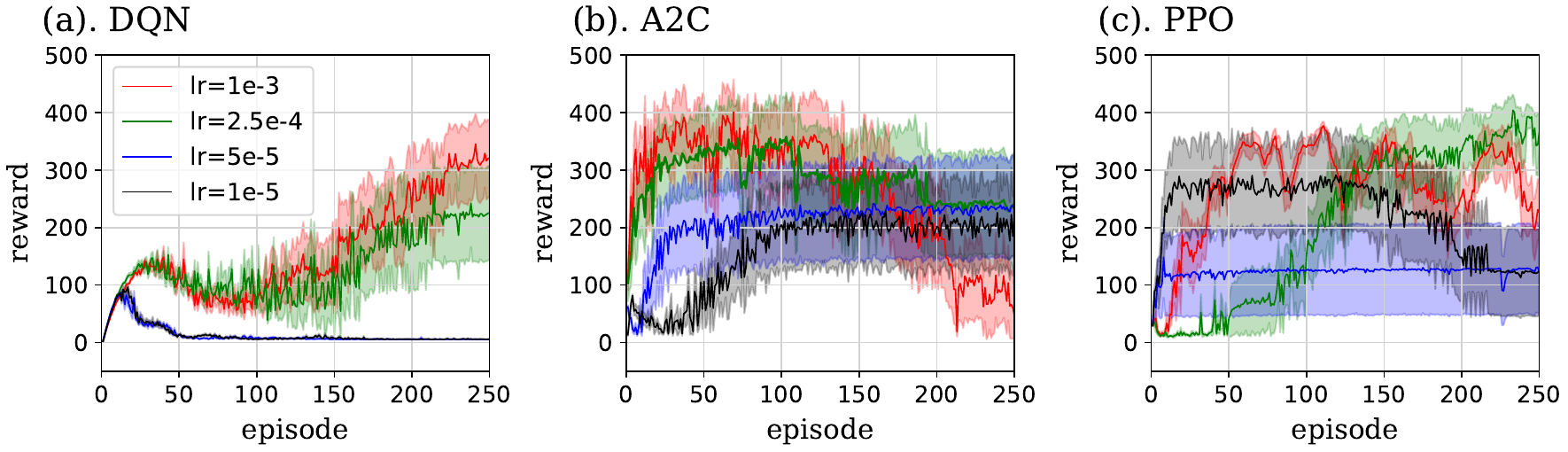}
\caption{Learning curves of (a).DQN, (b).A2C, and (c).PPO in task.~II at $n_{\mathrm{agent}} = 48$. The red, green, blue and black curves are the mean value of all agents' rewards when the learning rate is $1 \times 10^{-3}, 2.5 \times 10^{-4}$, $5 \times 10^{-5}$, and $1 \times 10^{-5}$, respectively. }
\label{LC_n48_lr}
\end{figure}

In task.~II with $n_{\mathrm{agent}} = 48$, the performance of DRL methods were worse than our method. It is because DRL methods failed to learn to make lanes and stagnated in some trials, whereas our method succeeded in learning in all trials. In this task, we investigated the performance of each DRL method for four values of the learning rate, $\eta = 1 \times 10^{-3}, 2.5 \times 10^{-4}$, $5 \times 10^{-5}$, and $1 \times 10^{-5}$, because it was possible that choice of appropriate hyperparameter led to improve the learning. The result is shown in Fig.~\ref{LC_n48_lr}. As we can see from this figure, stagnation happened in almost all learning rates of every method. Note that the case of PPO with $\eta = 1 \times 10^{-3}$ succeeded in learning to make lanes in every trial exceptionally, however, $\eta$ of this case was so large that the learning became unstable. In Fig.~\ref{LC_lane_DRL} (b), we chose the learning rate $\eta = 1 \times 10^{-3}$ for DQN (the red curve of Fig.~\ref{LC_n48_lr} (a)), and $\eta = 2.5 \times 10^{-4}$ for A2C and PPO (the green curves of Fig.~\ref{LC_n48_lr} (b) and (c)), because these values seemed to perform better than other values in each method. We actually executed the similar calculation for DQN in the case that minibatch size is 64, the same size used for other tasks. However, all trials for each of four learning rate $\eta$ failed to learn to make lanes in this case. Hence we increased the minibatch size to improve the sampling efficiency. The similar investigation was also executed for the case of task.~II with $n_{\mathrm{agent}} = 64$, i.e. we calculated the performance of each DRL method for four learning rates for this case. However, in this case, only PPO with $\eta = 1 \times 10^{-3}$ succeeded or began to secceed in making lanes in some trials, while the other methods and other learning rates failed in all trials.

It is difficult to interpret the result of task.~II because there are many unknown points on the algorithm combining ESN and LSPI method. One possibility is that agents choosing wrong actions cause the unclogging, as we discussed in Section \ref{1group}. Indeed, ESN has lower information processing capacity and choose the wrong action more frequently than DRL methods, as we saw in task.~I with $n_{\mathrm{agent}} = 40$. This point is thought to led to the unclogging and final success in learning under $n_{\mathrm{agent}} = 48$. In addition, PPO with $\eta = 1 \times 10^{-3}$, which succeeded in making lanes in all trials of $n_{\mathrm{agent}} = 48$ and some trials of $n_{\mathrm{agent}} = 64$, has unstable learning curves because of large learning rate, as we explained above. It is thought that such unstable behavior of agents also results in the unclogging. The other possible factor of the success of our method under $n_{\mathrm{agent}} = 48$ is that sampling efficiency of the LSPI method is higher than the gradient descent method, because it utilizes all experiences gathered by agents completely by the linear regression. However, our method failed in making lanes under $n_{\mathrm{agent}} = 64$, and the performance of this case was reversed by that of PPO with $\eta = 1 \times 10^{-3}$. Considering this point, effects of the above mentioned two factors on the final performance are thought to be complex. Hence, these effects should be studied also in other RL tasks we did not consider.  

We also measured the computational time of whole simulation in task.~I at $n_{\mathrm{agent}} = 1$ and 12 for each method, and summarize the result in Table~\ref{spent_time} (a). The calculation was executed by a laptop equipped with 12th Gen Intel(R) Core(TM) i9-12900H, and application software other than resident one was closed during the measurement of time. The real computational time for each simulation shown in Table~\ref{spent_time} (a) includes the time spent on the calculation not related to the neural network, such as updating of the environment. Hence, to evaluate the computational cost of the calculation on the neural network including both training and forward propagation, we calculated the computational time of the case when agents choose their action randomly, and showed the difference of each simulation from this case in Table~\ref{spent_time} (b). Here, in the random-action case, the neural network itself was not implemented. According to this table, computational cost of our method is less than half of that of each DRL method in both cases of $n_{\mathrm{agent}} = 1$ and 12. However, as for the difference of time between these two cases, which corresponds to increase in computational time due to increase in agents, DQN with its minibatch size 64 is better than other methods and our method is second to this. This is because the time spent for the calculation of backpropagation for DQN under fixed minibatch size does not depend on $n_{\mathrm{agent}}$, whereas minibatch size for A2C, replay-memory size for PPO, and the size of matrices used in our method, $Y_1, Y_2$, and $R$ given by eqs.~(\ref{Y_matrix}) and (\ref{R_matrix}), are all proportional to $n_{\mathrm{agent}}$. Considering the possibility that fixing minibatch size of DQN results in the worsening of sampling efficiency, we cannot say that DQN is more efficient algorithm than ours under a large number of agents, only from Table~\ref{spent_time} (b). Indeed, in the case of difficult tasks such as task.~II with $n_{\mathrm{agent}} = 48$, we should increase the minibatch size for DQN at the cost of the computational time, as we explained above. Hence, the computational cost of our method under large $n_{\mathrm{agent}}$ is lower than those of DRL methods, whose sampling efficiency do not reduce as $n_{\mathrm{agent}}$ increases.
\begin{table}
\centering
\begin{tabular}{cccc}
\multicolumn{4}{l}{(a). real computational time} \\
method & $n_{\mathrm{agent} } = 1$ & $n_{\mathrm{agent} } = 12$ & difference \\ \hline\hline
ESN+LSPI & \textbf{52.14(8)} & \textbf{420(7)} & 368(7) \\ \hline
DQN (minibatch size $=64$) & 715(7) & 898(5) & \textbf{183(9)} \\ 
DQN (minibatch size $=1024$) & 2645(4) & 4215(9) & 1570(10) \\ \hline
A2C & 202(2) & 1159(1) & 958(2) \\ \hline
PPO & 164(1) & 1301(21) & 1137(21) \\ \hline\hline
random action & 1.516(6) & 11.1(2) & 9.6(2) \\ \hline
\end{tabular}

\vspace{10.0mm}

\begin{tabular}{cccc}
\multicolumn{4}{l}{(b). difference from the random-action case} \\
method & $n_{\mathrm{agent} } = 1$ & $n_{\mathrm{agent} } = 12$ & difference \\ \hline\hline
ESN+LSPI & \textbf{50.62(8)} & \textbf{409(7)} & 358(7) \\ \hline
DQN (minibatch size $=64$) & 713(7) & 886(5) & \textbf{173(9)} \\ 
DQN (minibatch size $=1024$) & 2643(4) & 4204(9) & 1560(10) \\ \hline
A2C & 200(2) & 1148(1) & 948(2) \\ \hline
PPO & 163(1) & 1290(21) & 1127(21) \\ \hline
\end{tabular}
\vspace{1.0mm}
\caption{(a).Time spent for simulation using each method in task.~I with $n_{\mathrm{agent} } = 1$ and 12, averaged over 8 independent trials (measured in seconds). The right column indicates the difference between these two times. (b).Difference of the time from the case when the agents choose their action randomly. }
\label{spent_time}
\end{table}

\section{Summary \label{summary} }

In this study, we implemented MARL using ESN, and applied it to a simulation of pedestrian dynamics. The LSPI method was utilized to calculate the output weight matrix of the reservoir. As the environment, the grid-world composed of vacant space and walls was considered, and several agents were placed in it. Specifically, we investigated two types of tasks: I. a forked road composed of a narrow direct route and a broad detour, and II. a corridor where two groups of agents proceeded in the opposite directions. The simulations confirmed that the agents could learn to move forward by avoiding other agents, provided the density of the agents was not that high. 

In this study, we considered the case wherein the number of agents were at most 64. However, hundreds of agents are required for the simulations of real roads, intersections, or evacuation routes. Furthermore, to investigate certain phenomena such as the jamming transition mentioned in Section \ref{result_lane}, larger number of agents should be considered. Using deep learning, implementing such numerous RL agents may result in the enormous computational complexity. Hence, it is expected that reservoir computing including ESN, which can reflect agents' experiences efficiently with comparatively low computational cost, can be utilized to execute these studies in future. Note that RL-based methods including ours require time for training, compared with the traditional pedestrian simulations where agents obey the previously given rules. However, these methods have advantage that agents can automatically learn the appropriate actions without complex rule settings. In this study, for example, agents learned to avoid obstacles such as other agents without introducing repulsive forces or additional rewards for avoiding itself. Hence, the proposed method is thought to be useful in complex tasks such as the evacuation route flittered with obstacles. In addition, we believe that this method is applicable not only to the case of the pedestrian dynamics, but also to other behaviors of animal groups such as competition for food and territories.

Along with such applied studies, the improvement in the performance of reservoir computing itself is also important. Presently, the performance of reservoir computing tends to be poor under difficult tasks because of its low information processing capacity. For example, in image recognition, reservoir computing-based approaches exhibit low accuracy rates at CIFAR-10, although they succeed in recognizing simpler tasks such as MNIST\cite{ZV23,TT22}. In this context, in the case of the video games that use images as input data, reservoir computing-based RL has yielded only a few successful examples\cite{CF20}. In our study, we observed lower performance than DRL methods in some tasks, as we saw in Sec.~\ref{DRL}. In addition, considering that the learning speed under $\epsilon$-greedy method is limited by the value of $\epsilon$, we should study effective algorithms for exploration so that we can decrease this value fast.
 However, as we discussed in Sec.~\ref{DRL}, it is possible that our method has advantages not only in computational cost but also in other points such as sampling efficiency, compared to DRL ones. Hence, if the disadvantages mentioned above is improved in the future studies, MARL-based simulations of group behaviors of humans or animals will be further promoted.
 
In addition, whether the parameter sharing between different groups like Section \ref{1group} improve the sampling efficiency remains to be unknown. This point should also be studied in the future works, to improve the performance of MARL itself.
 
The source code for this work is uploaded to \url{https://github.com/Hisato-Komatsu/MARL_ESN_pedestrian}.


\section*{Acknowledgements}
We would like to thank Editage for English language editing.

\end{document}